\documentclass[12pt,english]{iopart}
\usepackage{graphicx}
\usepackage{dcolumn}
\usepackage{bm}
\usepackage[a4paper]{geometry}
\usepackage{units}
\usepackage{graphicx}
\usepackage{amssymb}
\usepackage{float}
\usepackage{setstack}

\makeatletter

\providecommand{\tabularnewline}{\\}

\newenvironment{lyxlist}[1]
{\begin{list}{}
{\settowidth{\labelwidth}{#1}
 \setlength{\leftmargin}{\labelwidth}
 \addtolength{\leftmargin}{\labelsep}
 }}
{\end{list}}

\usepackage{babel}
\usepackage{mathrsfs}
\usepackage{bm}

\newcommand{\ham}{\hat{\mathscr{H}}}
\newcommand{\cm}{\ensuremath{\hat{\mathscr{C}}}}

\date{}

\makeatother

\usepackage{babel}

\addtocounter{footnote}{-1}

\begin{document}

\title{Dynamics of a Bose-Einstein Condensate of Excited Magnons}

\author{F S Vannucchi, \'{A} R Vasconcellos\footnote{Deceased on October 13, 2012}\addtocounter{footnote}{5}, R Luzzi\footnote{Group Home Page: www.ifi.unicamp.br/$\sim$aurea}}

\address{Condensed Matter Physics Department, Institute of Physics {}``Gleb
Wataghin'', State University of Campinas - UNICAMP, 13083-859 Campinas,
SP, Brazil.}

\ead{fabiosv@ifi.unicamp.br}

\begin{abstract}
The emergence of a non-equilibrium Bose-Einstein-like condensation
of magnons in rf-pumped magnetic thin films has recently been experimentally
observed. We present here a complete theoretical description of the
non-equilibrium processes involved. It it demonstrated that the phenomenon
is another example of the presence of a Bose-Einstein-like condensation
in non-equilibrium many-boson systems embedded in a thermal bath, better 
referred-to as Fr\"{o}hlich-Bose-Einstein condensation.
The complex behavior emerges after a threshold of the exciting intensity
is attained. It is inhibited at higher intensities when the magnon-magnon
interaction drives the magnons to internal thermalization. The observed
behavior of the relaxation to equilibrium after the end of the pumping
pulse is also accounted for and the different processes fully described. 
\end{abstract}

\pacs{75.30.Ds, 05.70.Ln, 75.45.+j, 03.75.Nt}

\maketitle

\section{Introduction}

The kinetic of evolution of the system of spins in thin films of yttrium-iron-garnets
in the presence of a constant magnetic field, and being excited by
a source of rf-radiation which drives the system towards far-removed
from equilibrium conditions, has been reported in detailed experiments
performed by Demokritov et al. \cite{demokritov2006,demidov2008}.
These experimental results have evidenced the occurrence of an unexpected
large enhancement of the population of the magnons in the state lowest
in energy in their energy dispersion relation. That is, the energy
pumped on the system instead of being redistributed among the magnons
in such non-thermal conditions is transferred to the mode lowest in
frequency (with a fraction of course being dissipated to the surrounding
media). Some theoretical studies along certain approaches has been presented by several authors (see for example Refs. \cite{tupitsyn2008,rezende2009,malomed2010}); we proceed here to describe the phenomenon within a complete thermo-statistical description ensemble formalism. 

Such phenomenon has been referred-to as a non-equilibrium Bose-Einstein condensation, which then would belong to a family of three types of BEC:

The original one is the BEC in many-boson particle systems in equilibrium
at very low temperatures, which follows when their de Broglie thermal
wave length becomes larger than their mean separation distance, and
presenting some typical hallmarks (spontaneous symmetry breaking,
long-range coherence, etc.). Aside from the case of superfluidity,
BEC was realized in systems consisting of atomic alkali gases contained
in traps. A nice tutorial review is due to A. J. Leggett \cite{leggett2001}
(see also \cite{pitaevskii2003}).

A second type of BEC is the one of boson-like quasi-particles, that
is, those associated to elementary excitations in solids (e.g. phonons,
excitons, hybrid excitations, etc.), when in equilibrium at extremely
low temperatures. A well studied case is the one of an exciton-polariton
system confined in microcavities (a near two-dimensional sheet), exhibiting
the classic hallmarks of a BEC \cite{snoke2010}.

The third type, the one we are considering here, is the case of boson-like
quasi-particles (associated to elementary excitations in solids) which
are driven out of equilibrium by external perturbative sources. D.
Snoke \cite{snoke2006} has properly noticed that the name BEC can be
misleading (some authors call it {}``resonance'', e.g. in the case
of phonons \cite{phonon}), and following this author it is better
not to be haggling about names, and we introduce the nomenclature
NEFBEC (short for Non-Equilibrium Fr\"{o}hlich-Bose-Einstein Condensation
for the reasons stated below). As noticed, here we consider the case
of magnons (boson-like quasi-particles), demonstrating that NEFBEC
of magnons is another example of a phenomenon common to many-boson
systems embedded in a thermal bath (in the conditions that the interaction
of both generates non-linear processes) when driven sufficiently away
from equilibrium by the action of an external pumping source and which
display possible applications in the technologies of devices and medicine.
\begin{lyxlist}{00.00.0000}
\item [{1.}] A first case was evidenced by \textbf{Herbert Fr\"{o}hlich} who
considered the many boson system consisting of polar vibration (\textsc{lo}
phonons) in biopolymers under dark excitation (metabolic energy pumping)
and embedded in a surrounding fluid\cite{froehlich1970,froehlich1980,mesquita1993,fonseca2000}.
From a Science, Technology and Innovation (STI) point of view it was
considered to have implications in medical diagnosis\cite{hyland1998}.
More recently has been considered to be related to brain functioning
and artificial intelligence\cite{penrose1994}. 
\item [{2.}] A second case is the one of acoustic vibration (\textsc{ac}
phonons) in biological fluids, involving nonlinear anharmonic interactions
and in the presence of pumping sonic waves, with eventual STI relevance
in supersonic treatments and imaging in medicine\cite{lu1994,mesquita1998}. 
\item [{3.}] A third one is that of excitons (electron-hole pairs in semiconductors)
interacting with the lattice vibrations and under the action of rf-electromagnetic
fields; on a STI aspect, the phenomenon has been considered for allowing
a possible exciton-laser in the THz frequency range called {}``Excitoner''\cite{mysyrowics1996,mesquita2000}.
\item [{4.}] A fourth one is the case of magnons already referred to \cite{demokritov2006,demidov2008},
which we here analyze in depth. The thermal bath is constituted by
the phonon system, with which a nonlinear interaction exists, and
the magnons are driven arbitrarily out of equilibrium by a source of
electromagnetic radio frequency \cite{vannucchi2010}. Technological
applications are related to the construction of sources of coherent
microwave radiation \cite{demodov2011,ma2011}.
\end{lyxlist}
There exist two other cases of NEBEC (differing from NEFBEC) where
the phenomenon is associated to the action of the pumping procedure
of drifting electron excitation, namely,
\begin{lyxlist}{00.00.0000}
\item [{5.}] A fifth one consists in a system of longitudinal acoustic
phonons driven away from equilibrium by means of drifting electron
excitation (presence of an electric field producing an electron current),
which has been related to the creation of the so-called Saser, an
acoustic laser device, with applications in computing and imaging
\cite{kent2006,rodrigues2011}. 
\item [{6.}] A sixth one involving a system of LO-phonons driven away from
equilibrium by means of drifting electron excitation, which displays
a condensation in an off-center small region of the Brillouin Zone
\cite{rodrigues2010,komirenko2000}.
\end{lyxlist}
We describe here item number 4, namely, a system of magnons excited
by an external pumping source. For that purpose, we consider a system
of $N$ localized spins in the presence of a constant magnetic field,
being pumped by a rf-source of radiation driving them out of equilibrium
while embedded in a thermal bath consisting of the phonon system (the
lattice vibrations) in equilibrium with an external reservoir at temperature
$T_{0}$. The microscopic state of the system is characterized by
the full Hamiltonian of spins and lattice vibrations after going through
Holstein-Primakov and Bogoliubov transformations\cite{keffer1966,akhiezer1968,white1983}.
On the other hand, the characterization of the macroscopic state of
the magnon system is done in terms of the Thermo-Mechanical Statistics
based on the framework of a Non-Equilibrium Statistical Ensemble Formalism
(NESEF for short)\cite{luzzilivro2002,luzzi2006,zubarev1996,kuzensky2009,akhiezer1981,mclennan1963}.
Other modern approach consists in the use of Computational Modeling\cite{kalos2007,frenkel2002}
(developed after Non-equilibrium Molecular Dynamics\cite{alder1987}).
It may be noticed that NESEF is a systematization and an extension
of the essential contributions of several renowned authors following
the brilliant pioneering work of Ludwig Boltzmann. The formalism introduces
the fundamental properties of historicity and irreversibility in the
evolution of the nonequilibrium system where dissipative and pumping
processes are under way.

In terms of the dynamics generated by this full Hamiltonian the equations
of evolution of the macroscopic state of the system are obtained in
the framework of the NESEF-based nonlinear quantum kinetic theory
\cite{luzzilivro2002,luzzi2006,zubarev1996,kuzensky2009,akhiezer1981,mclennan1963,lauck1990,kuzensky2007,madureira1998,vannucchi2009osc}.
We call the attention to the fact that the evolution equations are
the quantum mechanical equations of motion averaged over the nonequilibrium
ensemble, with the NESEF-kinetic theory providing a practical way
of calculation. 

The paper is organized as follows:

In section 2 the Theoretical Background is described; In section 3
is presented the Evolution of the Nonequilibrium Macrostate of the
Magnon System; In section 4 NEFBEC in YIG is studied; In section 5
the Decay of the Condensate is analyzed; Finally, in section 6 we
present Additional Considerations and Concluding Remarks. In the Appendices
are included details of the derivation, which were omitted in the
main text in order to facilitate the reading.

\section{Theoretical Background}

We consider a system of localized effective spins characterized by
the Hamiltonian $\ham$

\begin{equation}
\ham=\ham_{\mathrm{exc}}+\ham_{\mathrm{dip}}+\ham_{\mathrm{Z}}+\ham_{\mathrm{SR}}+\ham_{\mathrm{R}}+\ham_{\mathrm{SL}}+\ham_{\mathrm{L}},\label{eq:hamiltoniana_total}\end{equation}
 where \begin{equation}
\ham_{\mathrm{exc}}=-\sum_{i,j\neq i}J(R_{ij})\mathbf{\hat{S}}_{i}\cdot\mathbf{\hat{S}}_{j},\label{eq:hamiltoniana_de_troca}\end{equation}
 accounts for the exchange interaction between pairs of localized
spins $\mathbf{\hat{S}}_{i}$ and $\mathbf{\hat{S}}_{j}$ in the equilibrium
positions $\mathbf{R}_{i}$ and $\mathbf{R}_{j}$ of the magnetic
ions that are present. With more than one per unit cell indexes $i$
and $j$ would be composed of the one indicating the position of the
unit cell and those indicating the positions of the magnetic ions
within the given unitary cell relative to the cell position. The exchange
integral $J$ depends on the distance $R_{ij}=\left|\mathbf{R}_{j}-\mathbf{R}_{i}\right|$.
The second contribution is the dipolar interaction given by 

\begin{equation}
\ham_{\mathrm{dip}}=\frac{(g\mu_{\mathrm{B}})^{2}}{2}\sum_{i,j\neq i}\left[\frac{\mathbf{\hat{S}}_{i}\cdot\mathbf{\hat{S}}_{j}}{R_{ij}^{3}}-\frac{3(\mathbf{\hat{S}}_{i}\cdot\mathbf{R}_{ij})(\mathbf{\hat{S}}_{j}\cdot\mathbf{R}_{ij})}{R_{ij}^{5}}\right]\label{eq:hamiltoniana_de_dipolo}\end{equation}
 with $g$ and $\mu_{\mathrm{B}}$ being the g-factor and Bohr's magneton
respectively. The terms $\ham_{\mathrm{Z}}$, $\ham_{\mathrm{SR}}$
and $\ham_{\mathrm{R}}$ are related with the magnetic field present in the
material: \begin{equation}
\ham_{\mathrm{Z}}=-g\mu_{\mathrm{B}}\mathbf{H}_{0}\cdot\sum_{i}\mathbf{\hat{S}}_{i},\label{eq:hamiltoniana_de_zeeman}\end{equation}
 is the so-called Zeeman term involving the coupling with an external
constant magnetic field $\mathbf{H}_{0}$; the time-dependent magnetic
fields $\mathbf{H}(t)$ (including the pumping rf-fields) are incorporated
through the term \begin{equation}
\ham_{\mathrm{SR}}=-g\mu_{\mathrm{B}}\mathbf{H}(t)\cdot\sum_{i}\mathbf{\hat{S}}_{i},\label{eq:hamiltoniana_spinluz}\end{equation}
 while $\ham_{\mathrm{R}}$ is the Hamiltonian of the free photons
in the electromagnetic fields.

The last terms account for the effects of lattice vibrations: $\ham_{\mathrm{L}}$
is the Hamiltonian of the free phonons and the spin-lattice interaction
is given by \cite{akhiezer1968} \begin{eqnarray}
\ham_{\mathrm{SL}}\:= & \sum_{i,j\neq i}\Biggl\{\left[\mathbf{x}_{ij}\cdot\frac{\partial}{\partial\mathbf{R}_{ij}}\right]\ham_{\mathrm{S}}(\mathbf{R}_{ij})+\nonumber \\
 & \qquad+\frac{1}{2}\left[\mathbf{x}_{ij}\cdot\frac{\partial}{\partial\mathbf{R}_{ij}}\right]^{2}\ham_{\mathrm{S}}(\mathbf{R}_{ij})\Biggr\},\label{eq:hamiltoniana_spinrede}\end{eqnarray}
with $\partial_{\mathbf{r}_{ij}}\ham_{\mathrm{S}}(\mathbf{R}_{ij})$
evaluated on the equilibrium positions $\mathbf{R}_{ij}$ and $\mathbf{x}_{ij}=\mathbf{x}_{j}-\mathbf{x}_{i}$
with $\mathbf{x}_{i}$ being the displacement of the ion around the
equilibrium position $\mathbf{R}_{i}$.

Introducing a second quantization formalism (which, in the case of
spins, is done in terms of Holstein-Primakov and Bogoliubov transformations
\cite{keffer1966,akhiezer1968,white1983}) we arrive to the expression
of the transformed Hamiltonian given in Appendix A {[}Eq. (\ref{eq:ham_linha_eq_cineticas}){]},
now in terms of magnon, phonon and photon creation (annihilation)
operators: $\hat{c}_{\mathbf{q}}^{\dagger}$ ($\hat{c}_{\mathbf{q}}$),
$\hat{b}_{\mathbf{q}}^{\dagger}$ ($\hat{b}_{\mathbf{q}}$) and $\hat{d}_{\mathbf{q}}^{\dagger}$
($\hat{d}_{\mathbf{q}}$) respectively. We call the Hamiltonian of
Eq. (\ref{eq:ham_linha_eq_cineticas}) the magnons' Hamiltonian, which
we use in the calculation of the nonequilibrium magnon populations.

For that purpose, first, the thermo-statistics deemed appropriate
for the description of the nonequilibrium macroscopic state of the
system needs be introduced. As noticed in the Introduction we resort
to the use of NESEF. According to the formalism, following Mori, Zwanzig
and others \cite{luzzilivro2002,luzzi2006,zubarev1996,kuzensky2009,akhiezer1981,mclennan1963},
the Hamiltonian of the system under consideration is separated out
in a so-called relevant part $\ham_{0}$, consisting of the energy
operators of the free degrees of freedom, and $\ham'$ containing
the interactions among them and the coupling with external sources
and reservoirs. In our case here \begin{eqnarray}
\ham_{0} & = \ham_{\mathrm{S}}^{(2)}+\ham_{\mathrm{L}}+\ham_{\mathrm{R}}=\nonumber \\
& = \sum_{\mathbf{q}}\hbar\omega_{\mathbf{q}}\hat{c}_{\mathbf{q}}^{\dagger}\hat{c}_{\mathbf{q}}+\sum_{\mathbf{k}}\hbar\Omega_{\mathbf{k}}\hat{b}_{\mathbf{k}}^{\dagger}\hat{b}_{\mathbf{k}}+\sum_{\mathbf{p}}\hbar\zeta_{\mathbf{p}}\hat{d}_{\mathbf{q}}^{\dagger}\hat{d}_{\mathbf{q}},\label{eq:ham_zero}\end{eqnarray}
 that is, the Hamiltonian of the free magnons {[}of Eq. (\ref{eq:hamiltoniana_magnonslivres}){]},
free phonons {[}only acoustic ones considered, Eq. (\ref{eq:hamiltoniano_fonons}){]}
and free photons introduced by the sample black-body radiation {[}Eq.
(\ref{eq:hamiltoniano_fotons}){]}, with quasi-momentum $\mathbf{q}$,
$\mathbf{k}$ and $\mathbf{p}$, and energy $\hbar\omega_{\mathbf{q}}$,
$\hbar\Omega_{\mathbf{k}}$ and $\hbar\zeta_{\mathbf{p}}$. On the
other hand, $\ham'$ contains the contributions associated to the
energies of interaction presented in Eqs. (\ref{eq:ham_linha_eq_cineticas}).

Next step in the application of the formalism consists in the choice
of a basic set of variables that should characterize the macroscopic
state of the system (the appropriate nonequilibrium thermodynamic
state of the system\cite{luzzi2000,luzzi2001}). At the microscopic
level, for the phonons and photons are taken the Hamiltonians $\ham_{\mathrm{L}}$
and $\ham_{\mathrm{R}}$, and for the magnon system is introduced
the free magnon Hamiltonian $\ham_{\mathrm{S}}^{(2)}$ and the magnetic
moment density operator \begin{equation}
\mathbf{\hat{M}}(\mathbf{r})=g\mu_{\mathrm{B}}\sum_{i}\mathbf{\hat{S}}_{i}\delta(\mathbf{r}-\mathbf{R}_{i}).\label{eq:M}\end{equation}

After applying Holstein-Primakoff and Bogoliubov transformations,
we do have that \begin{eqnarray}
\hat{\mathrm{M}}_{x}(\mathbf{r})& = g\mu_{\mathrm{B}}\sqrt{\frac{2S}{N}}\sum_{\mathbf{q}}\mbox{e}^{i\mathbf{q}\cdot\mathbf{r}}\left(v_{-\mathbf{q}}+u_{\mathbf{q}}\right)\left(\hat{c}_{-\mathbf{q}}^{\dagger}+\hat{c}_{\mathbf{q}}\right),\label{eq:Mx}\\
\hat{\mathrm{M}}_{y}(\mathbf{r})& = ig\mu_{\mathrm{B}}\sqrt{\frac{2S}{N}}\sum_{\mathbf{q}}\mbox{e}^{i\mathbf{q}\cdot\mathbf{r}}\left(v_{-\mathbf{q}}-u_{\mathbf{q}}\right)\left(\hat{c}_{-\mathbf{q}}^{\dagger}-\hat{c}_{\mathbf{q}}\right),\label{eq:My}\\
\hat{\mathrm{M}}_{z}(\mathbf{r})& = \mathrm{M}_{0}-\frac{g\mu_{\mathrm{B}}}{N}\sum_{\mathbf{q},\mathbf{q}'}\mbox{e}^{i(\mathbf{q}'-\mathbf{q})\cdot\mathbf{r}}\times\nonumber \\
 & \times\left\{ \left(u_{\mathbf{q}}^{*}u_{\mathbf{q}'}+v_{\mathbf{q}}v_{\mathbf{q}'}^{*}\right)c_{\mathbf{q}}^{\dagger}c_{\mathbf{q}'}+v_{\mathbf{q}}u_{\mathbf{q}'}c_{-\mathbf{q}}c_{\mathbf{q}'}+\right.\nonumber \\
 & \;\left.+u_{\mathbf{q}}^{*}v_{\mathbf{q}'}^{*}c_{\mathbf{q}}^{\dagger}c_{-\mathbf{q}'}^{\dagger}+v_{\mathbf{q}}v_{\mathbf{q}'}^{*}\delta_{\mathbf{q},\mathbf{q}'}\right\} ,\label{eq:Mz}\end{eqnarray}
 where in $\hat{\mathrm{M}}_{x}$ and $\hat{\mathrm{M}}_{y}$ have
been conserved only the linear contributions, $N$ is the number of
sites and we recall that $u_{\mathbf{q}}$ and $v_{\mathbf{q}}$ are
the coefficients in Bogoliubov transformation, given in Eq. (\ref{eq:coeficientes_bogoliubov})
in Appendix A.

Therefore, a priori, the set of basic microdynamical variables is
composed of \begin{equation}
\left\{ \ham_{\mathrm{S}}^{(2)},\,\mathbf{\hat{M}}(\mathbf{r}),\,\ham_{\mathrm{L}}+\ham_{\mathrm{R}}\right\} ,\label{eq:microdynamical_variables_1}\end{equation}
 that is, the free magnon Hamiltonian $\ham_{\mathrm{S}}^{(2)}$ of
Eq. (\ref{eq:hamiltoniana_magnonslivres}), the magnetization $\mathbf{\hat{M}}(\mathbf{r})$
of Eq. (\ref{eq:M}) and $\ham_{\mathrm{L}}+\ham_{\mathrm{R}}$ of
Eq. (\ref{eq:ham_zero}). However, for the purposes of the present
study, it is convenient to refine this basic set introducing each
of the contributions in $\ham_{\mathrm{S}}^{(2)}$ and in Eqs. \ref{eq:Mx},
\ref{eq:My} and \ref{eq:Mz}, namely \begin{eqnarray}
\Biggl\{\quad & \biggl\{\hat{\mathcal{N}}_{\mathbf{q}}=\hat{c}_{\mathbf{q}}^{\dagger}\hat{c}_{\mathbf{q}}\biggr\};\:\biggl\{\hat{\mathcal{N}}_{\mathbf{q},\mathbf{Q}}=\hat{c}_{\mathbf{q}+\frac{\mathbf{Q}}{2}}^{\dagger}\hat{c}_{\mathbf{q}-\frac{\mathbf{Q}}{2}}\biggr\};\:\nonumber \\
& \biggl\{\hat{\sigma}_{\mathbf{q}}^{\dagger}=\hat{c}_{-\mathbf{q}}^{\dagger}\hat{c}_{\mathbf{q}}^{\dagger}\biggr\};\:\biggl\{\hat{\sigma}_{\mathbf{q}}=\hat{c}_{-\mathbf{q}}\hat{c}_{\mathbf{q}}\biggr\};\nonumber \\
& \biggl\{\hat{\sigma}_{\mathbf{q},\mathbf{Q}}^{\dagger}=\hat{c}_{-\mathbf{q}-\frac{\mathbf{Q}}{2}}^{\dagger}\hat{c}_{\mathbf{q}-\frac{\mathbf{Q}}{2}}^{\dagger}\biggr\};\:\biggl\{\hat{\sigma}_{\mathbf{q},\mathbf{Q}}=\hat{c}_{-\mathbf{q}-\frac{\mathbf{Q}}{2}}\hat{c}_{\mathbf{q}-\frac{\mathbf{Q}}{2}}\biggr\};\nonumber \\
& \biggl\{\hat{c}_{\mathbf{q}}^{\dagger}\biggr\};\:\biggl\{\hat{c}_{\mathbf{q}}\biggr\};\:\ham_{\mathrm{L}}+\ham_{\mathrm{R}}\quad\Biggr\},\label{eq:microdynamical_variables_2}\end{eqnarray}
 where $\mathbf{Q}\neq0$.

It can be noticed that this set consists of components in reciprocal
space of the single-magnon reduced density matrix (Wigner - von Neumann
single-particle dynamical operator \cite{feynman1972,fano1957,balescu1975})
composed of the diagonal elements $\hat{N}_{\mathbf{q}}=\hat{c}_{\mathbf{q}}^{\dagger}\hat{c}_{\mathbf{q}}$,
the occupation number operators, and the non-diagonal ones $\hat{N}_{\mathbf{q},\mathbf{Q}}=\hat{c}_{\mathbf{q}+\mathbf{Q}/2}^{\dagger}\hat{c}_{\mathbf{q}-\mathbf{Q}/2}$
with $\mathbf{Q}\neq0$. The latter, describing the local inhomogeneities
of the occupations $\hat{N}_{\mathbf{q}}$, are not relevant for the
present problem once space-resolved experiments are not considered
at this point. It can be noticed that all the single-particle observables
of the system are expressed in terms of the single-particle reduced
density matrix \cite{fano1957}. Moreover, have been introduced the creation
($\hat{\sigma}_{\mathbf{q}}^{\dagger}$) and annihilation ($\hat{\sigma}_{\mathbf{q}}$)
operators of magnons pairs, and their nondiagonal contributions (associated
to local inhomogeneities) are not considered for the same reason appointed
above. Finally, being bosons, are introduced the creation and annihilation
operators in magnon states, $\hat{c}_{\mathbf{q}}^{\dagger}$ and
$\hat{c}_{\mathbf{q}}$, whose eigenstates are the so-called coherent
states. After this considerations
and recalling that $\ham_{\mathrm{S}}^{(2)}$ is expressed in terms
of occupation number operators {[}see Eq. (\ref{eq:hamiltoniana_magnonslivres}){]},
the set of microdynamical variables relevant to our problem is then
composed of \begin{equation}
\Biggl\{\biggl\{\hat{\mathcal{N}}_{\mathbf{q}}\biggr\};\:\biggl\{\hat{\sigma}_{\mathbf{q}}^{\dagger}\biggr\};\:\biggl\{\hat{\sigma}_{\mathbf{q}}\biggr\};\:\biggl\{\hat{c}_{\mathbf{q}}^{\dagger}\biggr\};\:\biggl\{\hat{c}_{\mathbf{q}}\biggr\};\:\ham_{\mathrm{L}}+\ham_{\mathrm{R}}\:\Biggr\}.\label{eq:microdynamical_variables_3}\end{equation}

According to NESEF, the nonequilibrium statistical operator, given
in Appendix B, depends on the quantities in set (\ref{eq:microdynamical_variables_3})
and on another set of nonequilibrium thermodynamic variables associated
(also said thermo-dynamically conjugated) to the basic ones in (\ref{eq:microdynamical_variables_3})
which we designate, respectively, by\begin{equation}
\Biggl\{\biggl\{ F_{\mathbf{q}}(t)\biggr\};\:\biggl\{\varphi_{\mathbf{q}}^{*}(t)\biggr\};\:\biggl\{\varphi_{\mathbf{q}}(t)\biggr\};\:\biggl\{\phi_{\mathbf{q}}^{*}(t)\biggr\};\:\biggl\{\phi_{\mathbf{q}}(t)\biggr\};\:\beta_{0}\:\Biggr\},\label{eq:thermovariables}\end{equation}
 with $\beta_{0}=1/k_{B}T_{0}$ {[}see Eq. (\ref{eq:rho_canonico}){]}.
Finally, the space of nonequilibrium thermodynamic variables consists
of the average values over the nonequilibrium ensemble of the quantities
in set (\ref{eq:microdynamical_variables_3}), say,
\small
\begin{equation}
\hspace{-1em}\Biggl\{\biggl\{\mathcal{N}_{\mathbf{q}}(t)\biggr\};\,\biggl\{\sigma_{\mathbf{q}}^{\dagger}(t)\biggr\};\,\biggl\{\sigma_{\mathbf{q}}(t)\biggr\};\,\biggl\{\left\langle \hat{c}_{\mathbf{q}}^{\dagger}|t\right\rangle \biggr\};\,\biggl\{\left\langle \hat{c}_{\mathbf{q}}|t\right\rangle \biggr\};\, E_{\mathrm{L}}+E_{\mathrm{R}}\,\Biggr\},\label{eq:macrovariables}\end{equation}
\normalsize
 that is, \begin{equation}
\mathcal{N}_{\mathbf{q}}(t)=\mbox{Tr}\left\{ \hat{\mathcal{N}}_{\mathbf{q}}\,\hat{\varrho}_{\varepsilon}(t)\times\hat{\varrho}_{\mathrm{B}}\right\}\end{equation}
 with $\hat{\varrho}_{\varepsilon}(t)$ of Eq. (\ref{eq:rho_final})
and $\hat{\varrho}_{\mathrm{B}}$ of Eq. (\ref{eq:rho_canonico}),
and so on for all the microdynamical variables in the set of Eq. (\ref{eq:microdynamical_variables_3}).

We are now in conditions to go over the derivation of the evolution
equations for the set (\ref{eq:macrovariables}) of basic macrovariables,
i.e. to obtain the time evolution of the nonequilibrium thermodynamic
state of the magnon system.

\section{Evolution of the Nonequilibrium Macrostate of the Magnon System\label{sec:Evolution}}

The equations of evolution for these variables are the quantum mechanical
equations of motion for the dynamical quantities of set (\ref{eq:microdynamical_variables_3})
averaged over the nonequilibrium ensemble. They are handled resorting
to the NESEF-based nonlinear quantum kinetic theory\cite{luzzilivro2002,lauck1990,kuzensky2007,madureira1998,vannucchi2009osc},
with the calculations performed in the approximation that incorporates
only terms quadratic in the interaction strength, with memory and
vertex renormalizations neglected\cite{lauck1990,madureira1998,zubarev?},
that is, we keep what in kinetic theory is called the irreducible
part of the two-particle collisions. In the case of the population
of magnons the equations of evolution are given by 
\begin{eqnarray}
\fl\frac{d}{dt}\mathcal{N}_{\mathbf{q}}(t) & = & \frac{1}{i\hbar}\mbox{Tr}\left\{ \left[\hat{\mathcal{N}}_{\mathbf{q}},\ham\right]\,\hat{\varrho}_{\varepsilon}(t)\times\hat{\varrho}_{\mathrm{B}}\right\} \simeq\nonumber \\
\fl & \simeq & \frac{1}{i\hbar}\mbox{Tr}\left\{ \left[\hat{\mathcal{N}}_{\mathbf{q}},\ham\right]\,\hat{\bar{\varrho}}(t,0)\times\hat{\varrho}_{\mathrm{B}}\right\} +\nonumber \\
\fl & & +\frac{1}{(i\hbar)^{2}}\int_{-\infty}^{t}dt'\mbox{ e}^{\varepsilon(t'-t)}\mbox{ Tr}\left\{ \left[\ham'(t'-t)_{0},[\ham',\hat{\mathcal{N}}_{\mathbf{q}}]\right]\,\hat{\bar{\varrho}}(t,0)\times\hat{\varrho}_{\mathrm{B}}\right\} +\nonumber \\
\fl & & +\frac{1}{i\hbar}\sum_{\ell}\int_{-\infty}^{t}dt'\mbox{ e}^{\varepsilon(t'-t)}\mbox{Tr}\left\{ [\ham'(t'-t)_{0},\hat{\mathcal{N}}_{\mathbf{q}}]\,\hat{\bar{\varrho}}(t,0)\times\hat{\varrho}_{\mathrm{B}}\right\} \frac{\delta J_{\mathcal{N}_{\mathbf{q}}}^{(1)}(t)}{\delta Q_{\ell}(t)},\label{eq:population_evolution_1}\end{eqnarray}
 where $\hat{\bar{\varrho}}(t,0)$ is an auxiliary statistical operator
(cf. Appendix B), $\varepsilon\to0$ after the calculation of averages,
lower index naught indicates interaction representation, $Q_{\ell}$
stands for the quantities in the set of Eq. (\ref{eq:macrovariables})
and $J_{\mathcal{N}_{\mathbf{q}}}^{(1)}=(i\hbar)^{-1}\mbox{Tr}\left\{ \left[\hat{\mathcal{N}}_{\mathbf{q}},\ham'\right]\,\hat{\bar{\varrho}}(t)\right\} $.
In the case of amplitudes, 

\begin{eqnarray}
\fl\frac{d}{dt}\left\langle \hat{c}_{\mathbf{q}}|t\right\rangle & = & \frac{1}{i\hbar}\mbox{Tr}\left\{ \left[\hat{c}_{\mathbf{q}},\ham\right]\,\hat{\varrho}_{\varepsilon}(t)\times\hat{\varrho}_{\mathrm{B}}\right\} \simeq\nonumber \\
\fl & \simeq & \frac{1}{i\hbar}\mbox{Tr}\left\{ \left[\hat{c}_{\mathbf{q}},\ham\right]\,\hat{\bar{\varrho}}(t,0)\times\hat{\varrho}_{\mathrm{B}}\right\} +\nonumber \\
\fl & & +\frac{1}{(i\hbar)^{2}}\int_{-\infty}^{t}dt'\mbox{ e}^{\varepsilon(t'-t)}\mbox{ Tr}\left\{ \left[\ham'(t'-t)_{0},[\ham',\hat{c}_{\mathbf{q}}]\right]\,\hat{\bar{\varrho}}(t,0)\times\hat{\varrho}_{\mathrm{B}}\right\} +\nonumber \\
\fl & & +\frac{1}{i\hbar}\sum_{\ell}\int_{-\infty}^{t}dt'\mbox{ e}^{\varepsilon(t'-t)}\mbox{Tr}\left\{ [\ham'(t'-t)_{0},\hat{c}_{\mathbf{q}}]\,\hat{\bar{\varrho}}(t,0)\times\hat{\varrho}_{\mathrm{B}}\right\} \frac{\delta J_{c_{\mathbf{q}}}^{(1)}(t)}{\delta Q_{\ell}(t)},\label{eq:amplitude_evolution_1}\end{eqnarray}
 the one for $\left\langle \hat{c}_{\mathbf{q}}^{\dagger}|t\right\rangle $
is the complex conjugated of this Eq. (\ref{eq:amplitude_evolution_1}),
 and the evolution equations for the magnon pairs are \begin{eqnarray}
\fl\frac{d}{dt}\sigma_{\mathbf{q}}(t)& = & \frac{1}{i\hbar}\mbox{Tr}\left\{ \left[\hat{\sigma}_{\mathbf{q}},\ham\right]\,\hat{\varrho}_{\varepsilon}(t)\times\hat{\varrho}_{\mathrm{B}}\right\} \simeq\nonumber \\
\fl& \simeq & \frac{1}{i\hbar}\mbox{Tr}\left\{ \left[\hat{\sigma}_{\mathbf{q}},\ham\right]\,\hat{\bar{\varrho}}(t,0)\times\hat{\varrho}_{\mathrm{B}}\right\} +\nonumber \\
\fl & & +\frac{1}{(i\hbar)^{2}}\int_{-\infty}^{t}dt'\mbox{ e}^{\varepsilon(t'-t)}\mbox{ Tr}\left\{ \left[\ham'(t'-t)_{0},[\ham',\hat{\sigma}_{\mathbf{q}}]\right]\,\hat{\bar{\varrho}}(t,0)\times\hat{\varrho}_{\mathrm{B}}\right\} +\nonumber \\
\fl & & +\frac{1}{i\hbar}\sum_{\ell}\int_{-\infty}^{t}dt'\mbox{ e}^{\varepsilon(t'-t)}\mbox{Tr}\left\{ [\ham'(t'-t)_{0},\hat{\sigma}_{\mathbf{q}}]\,\hat{\bar{\varrho}}(t,0)\times\hat{\varrho}_{\mathrm{B}}\right\} \frac{\delta J_{\sigma_{\mathbf{q}}}^{(1)}(t)}{\delta Q_{\ell}(t)},\label{eq:pairs_evolution_1}\end{eqnarray}
 and its complex conjugate for $\sigma_{\mathbf{q}}^{*}$.

They acquire the quite cumbersome expressions shown in Appendix C.
Here we present them in a compact form, indicating and describing
the contribution of the different processes involved, which for the
populations is\begin{equation}
\fl\frac{d}{dt}\mathcal{N}_{\mathbf{q}}(t)=\mathfrak{S}_{\mathbf{q}}(t)+\mathfrak{R}_{\mathbf{q}}(t)+L_{\mathbf{q}}(t)+\mathfrak{L}_{\mathbf{q}}(t)+\mathfrak{F}_{\mathbf{q}}(t)+\mathfrak{M}_{\mathbf{q}}(t)+\mathfrak{A}_{\mathbf{q}}(t)\,,\label{eq:population_evolution_2}\end{equation}

 where on the right we do have: (\emph{i}) $\mathfrak{S}_{\mathbf{q}}(t)$
is the rate of growth of the population in $\mathbf{q}$-mode produced
by the external source, which is composed of 2 contributions, namely,
a direct production, and a term of a positive feedback (only associated
to parallel pumping excitation); (\emph{ii}) $\mathfrak{R}_{\mathbf{q}}(t)$
is a non-linear term of relaxation due to the decay of magnon in photons,
leading to the saturation of absorption when under continuous excitation;
(\emph{iii}) $L_{\mathbf{q}}(t)$ is a term of linear relaxation to
the lattice with a relaxation time $\tau_{\mathbf{q}}$; (\emph{iv})
$\mathfrak{L}_{\mathbf{q}}(t)$ is a term involving nonlinear relaxation
to the lattice, referred to as Livshits contribution\cite{livshits1972};
(\emph{v}) $\mathfrak{F}_{\mathbf{q}}(t)$ is a peculiar and fundamental
contribution of a nonlinear character arising out of the magnon-lattice
interaction {[}the sixth and seventh contributions in Eq. (\ref{eq:N_q_evol})
of Appendix C{]}, which takes the form  

\begin{eqnarray}
\fl\mathfrak{F}_{\mathbf{q}}(t)& = {\displaystyle \frac{2\pi}{\hbar^{2}}\sum_{\mathbf{q}'\neq\mathbf{q}}}\left|\mathcal{F}_{\mathbf{q},\mathbf{q}-\mathbf{q}'}\right|^{2} & \left\{ \mathcal{N}_{\mathbf{q}'}(\mathcal{N}_{\mathbf{q}}+1)(\nu_{\mathbf{q}'-\mathbf{q}}+1)-(\mathcal{N}_{\mathbf{q}'}+1)\mathcal{N}_{\mathbf{q}}\nu_{\mathbf{q}'-\mathbf{q}}\right\}\times \nonumber \\
\fl & & \times\delta(\omega_{\mathbf{q}'}-\omega_{\mathbf{q}}-\Omega_{\mathbf{q}'-\mathbf{q}})+\nonumber \\
\fl & +{\displaystyle \frac{2\pi}{\hbar^{2}}\sum_{\mathbf{q}'\neq\mathbf{q}}}\left|\mathcal{F}_{\mathbf{q},\mathbf{q}-\mathbf{q}'}\right|^{2} & \left\{ (\mathcal{N}_{\mathbf{q}}+1)\mathcal{N}_{\mathbf{q}'}\nu_{\mathbf{q}-\mathbf{q}'}-\mathcal{N}_{\mathbf{q}}(\mathcal{N}_{\mathbf{q}'}+1)(\nu_{\mathbf{q}-\mathbf{q}'}+1)\right\}\times \nonumber \\
\fl & & \times\delta(\omega_{\mathbf{q}'}-\omega_{\mathbf{q}}+\Omega_{\mathbf{q}-\mathbf{q}'}),\label{eq:froehlich}\end{eqnarray}
where $\nu_{\mathbf{q}}$ and $\Omega_{\mathbf{q}}$ are the population
and the frequency dispersion relation of the phonons in the thermal
bath {[}see Eq. (\ref{eq:N_q_evol}) in Appendix C{]}. After some
mathematical handling, this Eq. (\ref{eq:froehlich}) can be rewritten
in the form\begin{equation}
\mathfrak{F}_{\mathbf{q}}(t)=\sum_{\mathbf{q}'}\chi_{\mathbf{qq}'}\left\{ \mathcal{N}_{\mathbf{q}'}(\mathcal{N}_{\mathbf{q}}+1)\,\mbox{e}^{\beta_{0}\hbar\omega_{\mathbf{q}'}}-(\mathcal{N}_{\mathbf{q}'}+1)\mathcal{N}_{\mathbf{q}}\,\mbox{e}^{\beta_{0}\hbar\omega_{\mathbf{q}}}\right\} ,\label{eq:froehlich2}\end{equation}
 where\begin{eqnarray}
\chi_{\mathbf{qq}'}=\frac{2\pi}{\hbar^{2}}\left|\mathcal{F}_{\mathbf{q},\mathbf{q}-\mathbf{q}'}\right|^{2}\big\{ & \nu_{\mathbf{q}-\mathbf{q}'}\,\mbox{e}^{-\beta_{0}\hbar\omega_{\mathbf{q}'}}\delta(\omega_{\mathbf{q}'}-\omega_{\mathbf{q}}+\Omega_{\mathbf{q}-\mathbf{q}'}) \nonumber \\
& +\nu_{\mathbf{q}'-\mathbf{q}}\,\mbox{e}^{-\beta_{0}\hbar\omega_{\mathbf{q}}}\delta(\omega_{\mathbf{q}'}-\omega_{\mathbf{q}}-\Omega_{\mathbf{q}-\mathbf{q}'})\big\} ,\label{eq:froehlich3}\end{eqnarray}
with this Eq. (\ref{eq:froehlich2}) having the form
given originally by Fr\"{o}hlich\cite{froehlich1970,froehlich1980}, and
we call it Fr\"{o}hlich contribution. 

Considering high population values ($\mathcal{N}_{\mathbf{q}'}\gg1$), Eq. (\ref{eq:froehlich2})
becomes \begin{equation}
\mathfrak{F}_{\mathbf{q}}(t)=\sum_{\mathbf{q}'}\chi_{\mathbf{qq}'}\mathcal{N}_{\mathbf{q}'}\mathcal{N}_{\mathbf{q}}\,(\mbox{e}^{\beta_{0}\hbar\omega_{\mathbf{q}'}}-\mbox{e}^{\beta_{0}\hbar\omega_{\mathbf{q}}}),\label{eq:froehlich4}
\end{equation} and we can see that, since $\chi_{\mathbf{qq}'}>0$, the contributions 
for the Eq. (\ref{eq:froehlich4}) are positive for those modes $\mathbf{q}^{\prime}$ 
for wich $\omega_{\mathbf{q}^{\prime}}>\omega_{\mathbf{q}}$ and negative 
for those modes with $\omega_{\mathbf{q}^{\prime}}<\omega_{\mathbf{q}}$. 
Consequently, modes $\mathbf{q}^{\prime}$ for
which $\omega_{\mathbf{q}^{\prime}}>\omega_{\mathbf{q}}$ transfer
their energy in excess of equilibrium to the mode $\mathbf{q}$, and
therefore in a cascade-down process it is transferred to the mode
lowest in frequency. Thus, the mode lowest in frequency largely grows
in population (drained from all the other modes) leading to the emergence
of, what has been dubbed, a nonequilibrium Bose-Einstein condensation.
Moreover, we emphasize that the Fr\"{o}hlich term has a purely quantum
mechanical origin. We can summarize the point stating that such \emph{nonequilibrium
Bose-Einstein condensation of {}``hot magnons'' is of a pure quantum
character and driven by Fr\"{o}hlich non-linear contribution to the kinetic
equations, whose origin is in the interaction with the thermal bath
in which the system is embedded} (a description of the irreversible
thermodynamics involved is presented in Ref. \cite{fonseca2000}). 

The other contributions in Eq. (\ref{eq:population_evolution_2})
are: (\emph{vi}) $\mathfrak{M}_{\mathbf{q}}(t)$ is the rate of change
generated by the magnon-magnon interaction (exchange and dipolar),
whose role is to lead the system of magnons to a state of nonequilibrium
internal thermalization. This contribution is in a {}``tug of war''
with Fr\"{o}hlich contribution (previous item); (\emph{vii}) $\mathfrak{A}_{\mathbf{q}}(t)$
contains all the contributions coupling the populations to the amplitudes,
$\left\langle \hat{c}_{\mathbf{q}}^{\dagger}|t\right\rangle $ and
$\left\langle \hat{c}_{\mathbf{q}}|t\right\rangle $, and the pair
functions, $\sigma_{\mathbf{q}}(t)$ and $\sigma_{\mathbf{q}}^{*}(t)$. 

Therefore this evolution equation is coupled to those of these other
basic variables given in Appendix C. The evolution equation for the
amplitudes, also in a compact form, is given by \begin{equation}
\frac{d}{dt}\left\langle \hat{c}_{\mathbf{q}}|t\right\rangle =-i\omega_{\mathbf{q}}\left\langle \hat{c}_{\mathbf{q}}|t\right\rangle -\Gamma_{\mathbf{q}}(t)\left\langle \hat{c}_{\mathbf{q}}|t\right\rangle ,\label{eq:amplitude_evolution_2}\end{equation}
 where, in Mori's terminology\cite{mori1965}, the first term on the
right is a precession term and the second, cf. Eq. (\ref{eq:gama}),
is in balance a relaxation (damping) term containing contributions
arising out of the magnon-phonon interaction (of linear, Livshits
and Fr\"{o}hlich type in the nomenclature already introduced), of interaction
with the radiation fields, and from the magnon-magnon interaction.

Similarly, for the pair magnon function follows {[}cf. Eq. (\ref{eq:pares_evol}){]}
that 

\begin{equation}
\frac{d}{dt}\sigma_{\mathbf{q}}(t)=-2i\omega_{\mathbf{q}}\,\sigma_{\mathbf{q}}-\Gamma_{\mathbf{q}}(t)\,\sigma_{\mathbf{q}}+\Lambda_{\mathbf{q}}(t),\label{eq:pairs_evolution_2}\end{equation}
 which is the first harmonic of the basic one in Eq. (\ref{eq:amplitude_evolution_2}),
and $\Lambda_{\mathbf{q}}(t)$ consists of a term involving the effects
of the external rf source and contributions coupling via interaction
with the latter and the magnon-magnon interaction, with the other
$\sigma_{\mathbf{q}'}(t)$ $\mathbf{q}'\neq\mathbf{q}$. Solution
of Eqs. (\ref{eq:population_evolution_2}), (\ref{eq:amplitude_evolution_2})
and (\ref{eq:pairs_evolution_2}) requires to provide initial conditions.
Considering the initial state as a equilibrium one, the initial condition
for populations $\mathcal{N}_{\mathbf{q}}(t)$ is $\mathcal{N}(t=0)=\left(\mbox{e}^{\beta_{0}\hbar\omega_{\mathbf{q}}}-1\right)^{-1}$,
and those for $\left\langle \hat{c}_{\mathbf{q}}|t=0\right\rangle $
and $\sigma_{\mathbf{q}}(t=0)$ are zero. Therefore, in the conditions
to be analyzed, $\left\langle \hat{c}_{\mathbf{q}}|t\right\rangle $
remains null (i.e., there is no contribution to the total population
from the population of the coherent states). On the other hand, $\left|\sigma_{\mathbf{q}}(t)\right|$
tends to increase with source $\Lambda_{\mathbf{q}}(t)$ (and to decay
with a lifetime given roughly by a time average of $\Gamma_{\mathbf{q}}^{-1}(t)$
during the length of the process), but the increase involves a rate
of change similar to the one for $\mathcal{N}_{\mathbf{q}}(t)$ and
then, in quite general conditions, the contribution to the population
due to the one of the magnon pairs is orders of magnitude smaller
than the leading one corresponding to that of the individual quasi-particles
(single magnons).

This can be seen in the fact that, on the one hand, Eq. (\ref{eq:amplitude_evolution_2})
can be expressed in the integral form \begin{equation}
\left\langle \hat{c}_{\mathbf{q}}|t\right\rangle =\left\langle \hat{c}_{\mathbf{q}}|0\right\rangle \exp\left\{ \int_{0}^{t}dt'\left[-i\omega_{\mathbf{q}}(t')-\Gamma_{\mathbf{q}}(t')\right]\right\} ,\end{equation}
 and, on the other hand, Eq. (\ref{eq:pairs_evolution_2}) becomes
\begin{eqnarray}
\sigma_{\mathbf{q}}(t)& = \sigma_{\mathbf{q}}(0)\exp\left\{ \int_{0}^{t}dt'\,\left[-i2\omega_{\mathbf{q}}(t)-\Gamma_{\mathbf{q}}(t')\right]\right\} +\nonumber \\
 & +\exp\left\{ \int_{0}^{t}dt'\,\left[-i2\omega_{\mathbf{q}}(t)-\Gamma_{\mathbf{q}}(t')\right]\right\} \times\nonumber \\
 & \times\int_{0}^{t}dt'\,\Lambda_{\mathbf{q}}(t')\exp\left\{ \int_{0}^{t'}dt''\,\left[i2\omega_{\mathbf{q}}(t)+\Gamma_{\mathbf{q}}(t'')\right]\right\} .\label{eq:pares_evol_integral}\end{eqnarray}

Moreover, a direct calculation provides us with the nonequilibrium
thermodynamic equations of state, that is, the relation between the
basic variables of the set (\ref{eq:macrovariables}) with the nonequilibrium
thermodynamic variables of set (\ref{eq:thermovariables}), namely
\begin{equation}
\mathcal{N}_{\mathbf{q}}(t)=\mbox{Tr}\left\{ \mathcal{N}_{\mathbf{q}}\,\hat{\varrho}_{\varepsilon}(t)\right\} =\mathcal{N}_{\mathbf{q}}^{\mathrm{sm}}(t)+\mathcal{N}_{\mathbf{q}}^{\mathrm{coh}}(t)+\mathcal{N}_{\mathbf{q}}^{\mathrm{pair}}(t),\end{equation}
 where we have introduced the definitions: \begin{equation}
\mathcal{N}_{\mathbf{q}}^{\mathrm{sm}}(t)\equiv\frac{1}{\mbox{e}^{F{}_{\mathbf{q}}}-1}\end{equation}
 is the population of the single magnons; \begin{equation}
\mathcal{N}_{\mathbf{q}}^{\mathrm{coh}}(t)\equiv\left|\left\langle \hat{c}_{\mathbf{q}}|t\right\rangle \right|^{2}=\left|\frac{\phi_{\mathbf{q}}(t)}{F_{\mathbf{q}}(t)+\left[\varphi_{\mathbf{q}}(t)+\varphi_{-\mathbf{q}}(t)\right]}\right|^{2}\end{equation}
 is the populations of the coherent states; and \begin{eqnarray}
\mathcal{N}_{\mathbf{q}}^{\mathrm{pair}}(t)\equiv\: & \frac{1}{\mbox{e}^{F{}_{\mathbf{q}}}-1}\left\{ \frac{F_{\mathbf{q}}(t)+F_{\mathbf{q}}'(t)}{2F_{\mathbf{q}}'(t)}\frac{\mbox{e}^{F{}_{\mathbf{q}}}-1}{\mbox{e}^{F'_{\mathbf{q}}}-1}+\right.\nonumber \\
 & \:\left.+\frac{F_{\mathbf{q}}(t)-F_{\mathbf{q}}'(t)}{2F_{\mathbf{q}}'(t)}\frac{\mbox{e}^{F{}_{\mathbf{q}}}-1}{1-\mbox{e}^{-F'_{-\mathbf{q}}}}-1\right\} \end{eqnarray}
 is the population of the magnon pairs; {\small $F_{\mathbf{q}}'(t)=\sqrt{F_{\mathbf{q}}^{2}(t)-\left|\varphi_{\mathbf{q}}(t)+\varphi_{-\mathbf{q}}(t)\right|^{2}}$}.
See appendix D for details. It can be noticed that we can redefine
the nonequilibrium thermodynamic variable $F_{\mathbf{q}}(t)$ in
either of two ways, \begin{equation}
F_{\mathbf{q}}(t)=\beta_{0}[\hbar\omega_{\mathbf{q}}-\mu_{\mathbf{q}}^{*}(t)]\mbox{ or }F_{\mathbf{q}}(t)=\hbar\omega_{\mathbf{q}}/k_{\mathrm{B}}T_{\mathbf{q}}^{\ast}(t)\,,\label{eq:quasi_thermovariables}\end{equation}
 following, respectively, Fr\"{o}hlich \cite{froehlich1970} and Landsberg
\cite{landsberg1981} introducing a so-called quasi-chemical potential
per mode, $\mu_{\mathbf{q}}^{*}(t)$, and Landau, Uhlenbeck and others
(a description in \cite{luzzilivro2002,luzzi1997}), introducing a
so-called quasi-temperature (or nonequilibrium temperature) per mode
\cite{luzzi1997,casas-vazquez2003}, $T_{\mathbf{q}}^{\ast}(t)$,
as it is usual in semiconductor physics \cite{kim1990,algarte1992}.

As described above, $\mathcal{N}_{\mathbf{q}}^{\mathrm{coh}}(t)=0$
and $\mathcal{N}_{\mathbf{q}}^{\mathrm{pair}}(t)\ll\mathcal{N}_{\mathbf{q}}^{\mathrm{sm}}(t)$,
and then in Eq. \ref{eq:population_evolution_2} we can take $\mathfrak{A}_{\mathbf{q}}(t)\simeq0$
(i.e., $\mathfrak{A}_{\mathbf{q}}(t)$ orders of magnitude smaller
than the other terms) and then Eq. \ref{eq:population_evolution_2}
is closed in itself and we proceed to deal with it.

\section{BEC in YIG}

\begin{figure}[H]
\begin{centering}
\includegraphics[width=0.5\columnwidth]{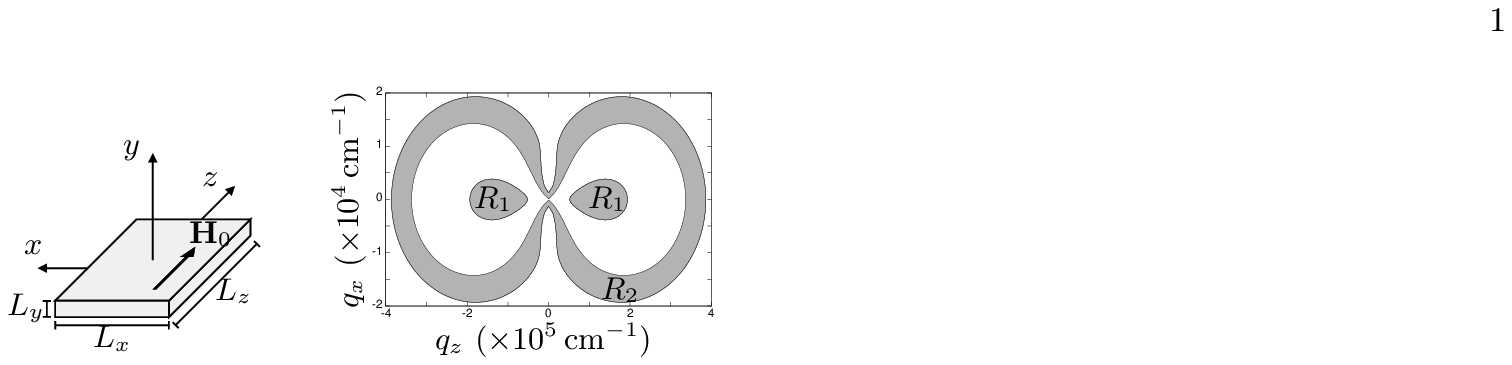}
\par\end{centering}

\caption{\label{fig:filme_fino+regioes}On left we see the thin film geometry,
with magnons propagating in the $xz$-plane. On right, the regions
$R_{1}$ and $R_{2}$ in quasimomentum space.}

\end{figure}

For numerical calculations and comparison with experiment, first we
introduce a treatment consisting in a kind of {}``two-fluid model'',
namely, we transform the large system of coupled evolution equations
in a pair of coupled equations for the mean values of the populations
over two regions of the Brillouin zone: one is a small region around
the position of the minimum in frequency in the dispersion relation
($\simeq\unit[2,1]{GHz}$), which we call $R_{1}$, and the other
around the zone in which are the modes pumped by the rf-fields ($\simeq\unit[4]{GHz}$),
indicated by $R_{2}$. Such kind of procedure is justified, first, because
of what followed in the other cases of BEC in bosons (phonons, excitons)
where a complete solution is obtained (symmetry conditions allowed
to separate the set of coupled equations in blocks with a small number
of coupled equations) showing that kind of behaviour and, second,
it is verified to a good degree in the experiments, as can be noticed
in Fig. 3 of Ref. \cite{demokritov2006} and better in Fig. 2 of Ref.
\cite{demidov2008}. The chosen regions are shown on the right of
Fig. \ref{fig:filme_fino+regioes} and were determined on the basis of 
what is shown by the experimental Brillouin spectra reported in Refs. \cite{demokritov2006} and \cite{demidov2008}.

\subsection{Evolution of the {}``Two Fluid Model''\label{subsec:BEC-in-YIG}}

In terms of the previous considerations we introduce the mean populations 
$\mathcal{N}_{1}(t)$ and $\mathcal{N}_{2}(t)$ corresponding to the relevant 
regions  $R_{1}$ and $R_{2}$ given by \begin{equation}
\mathcal{N}_{1}(t)=\frac{1}{n_{1}}\sum_{\mathbf{q}\in R_{1}}\mathcal{N}_{\mathbf{q}}(t),\end{equation}
\begin{equation}
\mathcal{N}_{2}(t)=\frac{1}{n_{2}}\sum_{\mathbf{q}\in R_{2}}\mathcal{N}_{\mathbf{q}}(t),\end{equation}
 where $n_{1,2}={\displaystyle \sum_{\mathbf{q}\in R_{1,2}}1}$ represents
the number of modes in the regions $R_{1}$ and $R_{2}$.

Proceeding correspondingly in Eq. \ref{eq:population_evolution_2},
we do have that (omitting to explicitly write the time
dependence on the right) \begin{eqnarray}
f_{1}{\displaystyle \frac{d}{d\bar{t}}}\mathcal{N}_{1}(\bar{t})& = -\mathrm{D}\,\mathcal{N}_{1}(\mathcal{N}_{1}-\mathcal{N}_{1}^{(0)})-f_{1}\left[\mathcal{N}_{1}-\mathcal{N}_{1}^{(0)}\right]+\nonumber \\
 & +\mathrm{F}\left\{ \mathcal{N}_{1}\mathcal{N}_{2}+\left(\bar{\nu}+1\right)\mathcal{N}_{2}-\bar{\nu}\mathcal{N}_{1}\right\} -\nonumber \\
 & -\mathrm{M}\left\{ \mathcal{N}_{1}\left(\mathcal{N}_{1}+1\right)+\mathcal{N}_{2}\left(\mathcal{N}_{2}+1\right)\right\} (\mathcal{N}_{1}\frac{\mathcal{N}_{2}^{(0)}}{\mathcal{N}_{1}^{(0)}}-\mathcal{N}_{2}),\label{eq:N1_evolution}\end{eqnarray}
 and \begin{eqnarray}
f_{2}{\displaystyle \frac{d}{d\bar{t}}}\mathcal{N}_{2}(\bar{t})& = \mathrm{I}\,(1+2\mathcal{N}_{2})-\nonumber \\
 & -\mathrm{D}\,\mathcal{N}_{2}(\mathcal{N}_{2}-\mathcal{N}_{2}^{(0)})-f_{2}\left[\mathcal{N}_{2}-\mathcal{N}_{2}^{(0)}\right]-\nonumber \\
 & -\mathrm{F}\left\{ \mathcal{N}_{1}\mathcal{N}_{2}+\left(\bar{\nu}+1\right)\mathcal{N}_{2}-\bar{\nu}\mathcal{N}_{1}\right\} +\nonumber \\
 & +\mathrm{M}\left\{ \mathcal{N}_{1}\left(\mathcal{N}_{1}+1\right)+\mathcal{N}_{2}\left(\mathcal{N}_{2}+1\right)\right\} (\mathcal{N}_{1}\frac{\mathcal{N}_{2}^{(0)}}{\mathcal{N}_{1}^{(0)}}-\mathcal{N}_{2}),\label{eq:N2_evolution}\end{eqnarray}
where $\bar{t}$ is the scaled time $t/\tau$, taking
the relaxation time $\tau_{\mathbf{q}}$ as having a unique constant
value ($\mathbf{q}$-independent), $\mathcal{N}_{1,2}^{(0)}$ are
the populations in equilibrium, and $f_{1}$ and $f_{2}$ the fractions
of the Brillouin zone corresponding to the two regions in the two-fluid
model. Moreover, the coefficients $\mathrm{M}$ and $\mathrm{F}$
are the coupling strengths associated to magnon-magnon interaction
and to Fr\"{o}hlich contribution respectively, $\mathrm{D}$ is the one
associated to decay with emission of photons, and $\bar{\nu}$ is
an average population of the phonons; the Livshits term can be neglected.
Finally, the parameter $\mathrm{I}$ is related to the rate of the
rf-radiation field transferred to the spin system, whose absorption,
as noticed, is reinforced by a positive feedback effect. All these coefficients
are addimensional, being multiplied by the relaxation time $\tau$.

Consider now the experiment in Ref. \cite{demokritov2006}. The fractions
$f_{1}$ and $f_{2}$ follows considering the size of regions $R_{1}$
and $R_{2}$, and, given the frequencies associated with these regions,
the values of the mean populations in equilibrium are $\mathcal{N}_{1}^{(0)}=3\times10^{3}$
and $\mathcal{N}_{2}^{(0)}=2\times10^{3}$. An effective absorbed
power of $\unit[10^{-2}]{W}$ would correspond to $\mathrm{I}\approx10^{-3}$.
On the other hand, an estimate of the $\mathcal{N}_{2}$ steady state
value allows to evaluate that $\mathrm{D}\approx10^{-11}$. On the
basis of a transient time for attaining internal thermalization near
equilibrium of the order of $\unit[200]{ns}$, it can be estimated
that $\mathrm{M}\lesssim10^{-12}$. Being $\tau$ of the order of
a few microseconds (and adopting $\tau=\unit[1]{\mu s}$), we are
left with $\mathrm{F}$ as the only open parameter.

\begin{figure}[H]
\begin{centering}
\includegraphics[width=0.5\columnwidth]{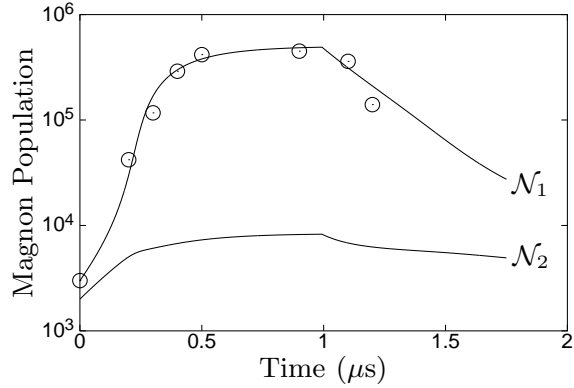}
\par\end{centering}

\caption{\label{fig:time} Evolution of the magnon population. Circles represent
Demokritov's et al. data for the low energy magnon population \cite{demokritov2006},
with the pumping being switched off after $\unit[1]{\mu s}$. Solid
lines show magnon populations of low and high energy, obtained after
numerical integration of Eqs. (\ref{eq:N1_evolution}) and (\ref{eq:N2_evolution})
using the following parameters: $f_{1}=3\times10^{-6}$; $f_{2}=3\times10^{-4}$;
$\mathcal{N}_{1}^{(0)}=3\times10^{3}$; $\mathcal{N}_{2}^{(0)}=2\times10^{3}$;
$\mathrm{I}=8\times10^{-4}$; $\mathrm{D}=4\times10^{-11}$; $\mathrm{M}=3\times10^{-14}$;
$\mathrm{F}=2\times10^{-6}$.}

\end{figure}

Using the experimental data \cite{demokritov2006}, varying the values
of the parameters around those given above and adjusting $\mathrm{F}$,
it follows the good agreement of theory and experiment shown in Fig.
\ref{fig:time}.

\begin{figure}[h]
\begin{centering}
\includegraphics[width=0.5\columnwidth]{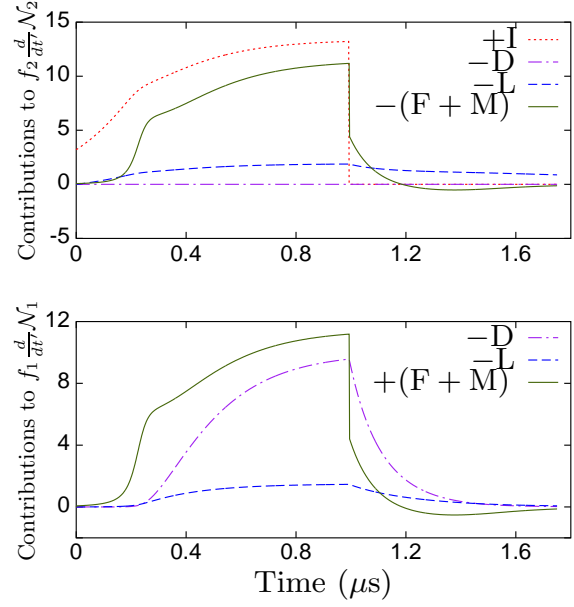} 
\par\end{centering}

\caption{\label{fig:time_processes} (Color online) Contributions (scaled) from all the distinct
processes to the evolution of $\mathcal{N}_{1}$ and $\mathcal{N}_{2}$.
With the same parameters used in Fig. \ref{fig:time}, we plotted
the rates of decay ($-$) and increase ($+$) of $\mathcal{N}_{2}$
(top) and $\mathcal{N}_{1}$ (bottom). }

\end{figure}

We proceed to analyse the several processes leading to the increase
of the populations in Fig. \ref{fig:time}, that are described in
Fig. \ref{fig:time_processes} in terms of the rates of increase and
decay, the contributions present on the right of equations (\ref{eq:N1_evolution})
and (\ref{eq:N2_evolution}). Multiplying these rates by the total
number of magnon modes $n\simeq3\times10^{14}$, we obtain the rates
related to the total number of magnons in regions $R_{1}$ and $R_{2}$.

Besides the rate of pumping from the external source (designated by
$\mathrm{I}$), are present the contribution arising out from Fr\"{o}hlich
effect, which is a pumping term for $\mathcal{N}_{1}$ and a decay
one for $\mathcal{N}_{2}$, and the contribution due to magnon-magnon
interaction, which redistributes the pumped energy among the modes,
tending to drive the system to nonequilibrium internal thermalization 
(non-scattering mechanisms, i.e., decay and emission are discarded for not 
being of relevance: energy and momentum conservation are impaired\cite{keffer1966}).
These two effects (whose composition is referred as $\mathrm{F}+\mathrm{M}$)
are responsible for internal interaction of magnons $\mathcal{N}_{1}$
and $\mathcal{N}_{2}$. Two other contributions correspond to decay,
namely, a linear term of decay to the lattice ($\mathrm{L}$) and
the bilinear one of decay by photon emission ($\mathrm{D}$). The
sign ($+$) for pumping and ($-$) for decay is indicated in the inset. 

During the time interval when the pumping is on, the source creates
magnons on region $R_{2}$, while internal interactions (Fr\"{o}hlich
and magnon-magnon) annihilate them. These internal interactions are
responsible for creating magnons in region $R_{1}$, that decay mainly
through photon emission. Thus, the nonequilibrium two-fluid system
has the energy pumped on region $R_{2}$, transferred to region $R_{1}$
via the composition of Fr\"{o}hlich and magnon-magnon effects and while
being lost through photon emission. 

The interplay of all these processes changes as the rate of the pumping
source is changed. The results reported above are for the scaled rate
$\mathrm{I}=\unit[8\times10^{-4}]{\mu\mbox{s}^{-1}}$.

\subsection{The Steady State}

\begin{figure}[h]
\begin{centering}
\includegraphics[width=0.5\columnwidth]{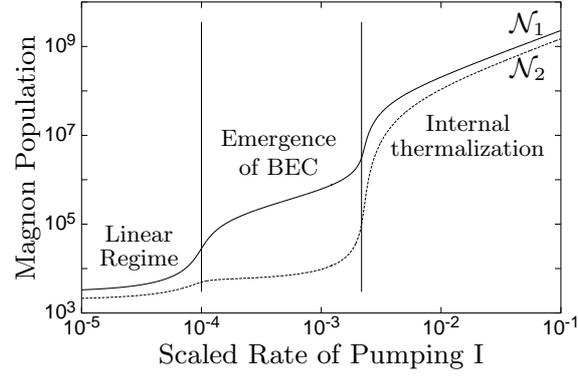}
\par\end{centering}

\caption{\label{fig:stationary}Steady-state magnon populations {[}solutions
of Eqs. (\ref{eq:N1_evolution}) and (\ref{eq:N2_evolution}){]} as
a function of the scaled rate of pumping due to the source, using
the same parameters as in Figure \ref{fig:time}. It can be noticed
the existence of a window for the emergence of BEC, which follows
at a certain threshold of intensity. In the linear regime, at low
intensity, no particular complex behaviour follows, and at high levels
of intensity magnon-magnon interaction overcomes Fr\"{o}hlich contribution
and there follows internal thermalization.}

\end{figure}

Using the same parameters but considering constant application of
the pumping source, the steady-state populations are obtained as a
function of the source scaled rate of pumping, what is shown in Fig.
\ref{fig:stationary}. It can be noticed the existence of a critical
pumping scaled rate (better saying, a rate threshold) after which
there follows a steep increase in the population of the mode lowest
in frequency, characterized by $\mathcal{N}_{1}$, corresponding to
the emergence of BEC. With increasing pumping intensity a second critical
rate (rate threshold) is evidenced such that for higher values of
$\mathrm{I}$ is observed internal thermalization of the magnons which
acquire a common quasi-temperature {[}cf. Eq. (\ref{eq:quasi_thermovariables}){]}.
This implies that the magnon-magnon interaction overcomes Fr\"{o}hlich
contribution and BEC is impaired.

\begin{figure}[H]
\begin{centering}
\includegraphics[width=0.5\columnwidth]{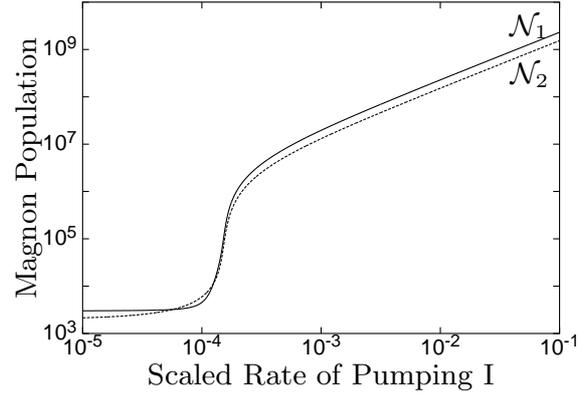} 
\par\end{centering}

\caption{\label{fig:stationary_F=00003D0}Steady-state magnon populations as
a function of the scaled rate of pumping when Fr\"{o}hlich term is disregarded
($\mathrm{F}=0$).}

\end{figure}

\begin{figure}[H]
\begin{centering}
\includegraphics[width=0.5\columnwidth]{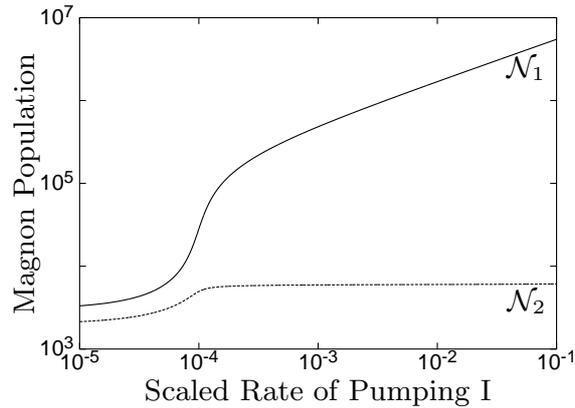} 
\par\end{centering}

\caption{\label{fig:stationary_M=00003D0}Steady-state magnon populations as
a function of the scaled rate of pumping when magnon-magnon interaction
is disregarded ($\mathrm{M}=0$).}

\end{figure}

We can better appreciate the role of both types of interactions in
figures \ref{fig:stationary_F=00003D0} and \ref{fig:stationary_M=00003D0}.
In Fig. \ref{fig:stationary_F=00003D0} the Fr\"{o}hlich contribution
is {}``switched off'' ($\mathrm{F}=0$), and the action of the magnon-magnon
leading to internal thermalization (for $\mathrm{I}>3\times10^{-4}$)
is evidenced. 

On the contrary, in Fig. \ref{fig:stationary_M=00003D0} where the
magnon-magnon interaction is {}``switched off'' ($\mathrm{M}=0$),
the emergence of NEFBEC (for $\mathrm{I}>10^{-4}$) does follow unimpeded
by the magnon-magnon interaction.

\begin{figure}[h]
\begin{centering}
\includegraphics[width=0.5\columnwidth]{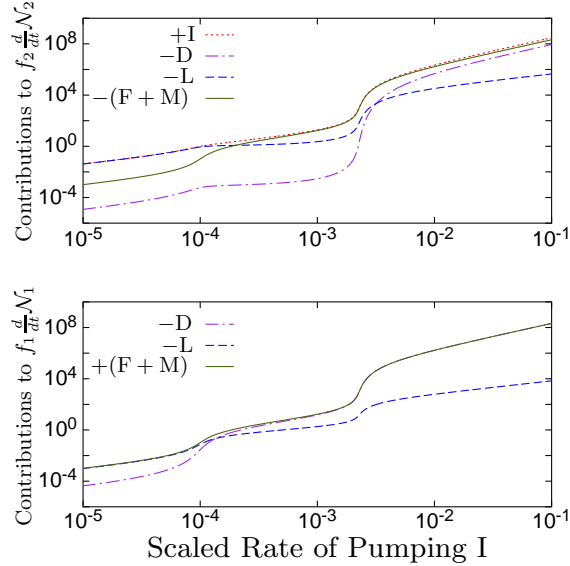}
\par\end{centering}

\caption{\label{fig:processes_stationary} (Color online) Role of contributions (scaled) from
all the distinct processes to the stationary state of $\mathcal{N}_{1}$
and $\mathcal{N}_{2}$. The rates for $\mathcal{N}_{2}$ (top) and
$\mathcal{N}_{1}$ (bottom), with the same notation of Figure \ref{fig:time_processes},
are presented. }

\end{figure}

Returning to Fig. \ref{fig:stationary}, let us consider the interplay
of the several mechanisms that lead to the formation of the steady
state, when the rates of change are balanced. The rate of change associated
with all the mentioned processes, in the steady state, for a range
of values of the (scaled) rate of pumping from the external source,
are shown in figure \ref{fig:processes_stationary}, in the upper
part for $\mathcal{N}_{2}$ and in the lower part for $\mathcal{N}_{1}$,
using the same notation of Fig. \ref{fig:time_processes}. It can
be noticed the presence of three regimes in correspondence with those
indicated in Fig. \ref{fig:stationary}. In all cases energy is fed
to the system through the {}``$\mathcal{N}_{2}$ magnons'' pumped
by the source, while Fr\"{o}hlich and magnon-magnon redistribute it, creating
{}``$\mathcal{N}_{1}$ magnons''. In the range of scaled pumping
rates up to roughly $6\times10^{-5}$, the linear regime, both $\mathcal{N}_{1}$
and $\mathcal{N}_{2}$ magnons are relaxing predominantly through
the linear decay to the lattice. For greater intensities the linear
decay to the lattice is not sufficient to wholly absorb the system
energy in order to maintain the steady state. Thus, it is observed
an increase on the $\mathcal{N}_{1}$ populations, characterizing
the condensate, and in order to maintain the system stationary other
relaxation mechanisms become predominant: {}``$\mathcal{N}_{1}$
magnons'' decay through photon emission and {}``$\mathcal{N}_{2}$
magnons'' through the $\mathrm{F}+\mathrm{M}$ term. Finally, for
$\mathrm{I}\gtrsim2\times10^{-3}$, the non-linear photon emission
decay gains relevance, the $\mathcal{N}_{2}$ populations increase,
and the internal thermalization regime is attained.

All this interplay explains, in a sense, the transition between the
three regimes: In the linear regime, as the rate of energy transfer
from the source increases, we have that, at low scaled rates of pumping,
the populations attain values near the equilibrium ones, the usual
linear relaxation to the lattice, with a characteristic relaxation
time, predominates and no particular complex behavior is to be expected.
As the intensity approaches a threshold value, the nonlinear contributions,
associated to Fr\"{o}hlich effect, magnon-magnon interactions and photon
emission, begin to be relevant as the populations increase to values
much greater than in equilibrium. Fr\"{o}hlich effect leads to the emergence
of BEC once it overcomes the tendency for internal thermalization
promoted by the magnon-magnon interaction, and the decay by photon
emission ensures that a steady state is attained. The second threshold
in intensity value follows as magnon-magnon interaction overcomes
Fr\"{o}hlich effect and internal thermalization is ensured.

\begin{figure}[H]
\begin{centering}
\includegraphics[width=0.5\columnwidth]{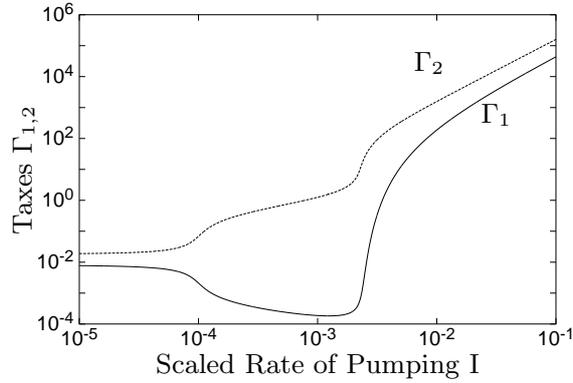} 
\par\end{centering}

\caption{\label{fig:stationary_gama}Values of $\Gamma_{1,2}$ in the steady
state (cf. Eqs. \ref{eq:gamma1} e \ref{eq:gamma2}) associated with
the magnon populations (Figure \ref{fig:stationary}).}

\end{figure}

Moreover, considering the quantity $\Gamma_{\mathbf{q}}$ of Eqs.
\ref{eq:amplitude_evolution_2} and \ref{eq:pairs_evolution_2}, in
this two-fluid model we do have for $\Gamma_{1}$ and $\Gamma_{2}$
(mean values of $\Gamma_{\mathbf{q}}$ in regions $R_{1}$ and $R_{2}$)
that \begin{eqnarray}
\fl\Gamma_{1}(\bar{t}) = & \mathrm{D}\,(\mathcal{N}_{1}-\mathcal{N}_{1}^{(0)})+f_{1}-\mathrm{F}\,(\mathcal{N}_{2}-\bar{\nu})+\nonumber \\
\fl & +\mathrm{M}\left\{ \frac{\mathcal{N}_{2}^{(0)}}{\mathcal{N}_{1}^{(0)}}\left[\mathcal{N}_{1}\left(\mathcal{N}_{1}+1\right)+\mathcal{N}_{2}\left(\mathcal{N}_{2}+1\right)\right]-\left(\mathcal{N}_{1}+1\right)\mathcal{N}_{2}\right\} ,\label{eq:gamma1}\end{eqnarray}
 and \begin{eqnarray}
\fl\Gamma_{2}(\bar{t})= & -2\mathrm{I}+\mathrm{D}\,(\mathcal{N}_{2}-\mathcal{N}_{2}^{(0)})+f_{2}+\mathrm{F}\,\left(\mathcal{N}_{1}+\bar{\nu}+1\right)+\nonumber \\
\fl & \vspace{-3cm}+\mathrm{M}\left\{ \mathcal{N}\left(\mathcal{N}_{1}+1\right)+\mathcal{N}_{2}\left(\mathcal{N}_{2}+1\right)-\frac{\mathcal{N}_{2}^{(0)}}{\mathcal{N}_{1}^{(0)}}\mathcal{N}_{1}\left(\mathcal{N}_{2}+1\right)\right\} ,\label{eq:gamma2}\end{eqnarray}
 which have their stationary values shown in Fig. \ref{fig:stationary_gama}
as a function of the rate of pumping $\mathrm{I}$.

According to the considerations presented in section \ref{sec:Evolution},
we see that $\Gamma_{\mathbf{q}}$ is the linear decay rate of the
amplitudes, $\left\langle \hat{c}_{\mathbf{q}}^{\dagger}|t\right\rangle $
and $\left\langle \hat{c}_{\mathbf{q}}|t\right\rangle $. All the
obtained values of $\Gamma_{1,2}$ are positive, (Fig. \ref{fig:stationary_gama}),
indicating decay of these amplitudes, which corroborates the assumed
neglect of their contribution to the populations' evolution. Although
the amplitudes decay, we point that in the source intensity interval
associated with the condensate, $10^{-4}\lesssim\mathrm{I}\lesssim2\times10^{-3}$,
$\Gamma_{1}$ is considerably smaller than $\Gamma_{2}$, and the
condensate would allow long mean-life to the coherent states corresponding
to the lowest frequency modes, which can be excited to compose solitary
waves, as it has been shown in the case of other boson systems displaying
NEFBEC \cite{fonseca2000,mesquita1998,mesquita2000}.

The decay of the populations after the switching off of the pumping
source follows in accord with the
one observed in the experiment of Ref. \cite{demidov2008b}. It consists
of three regimes: a near exponential one at the initial delay times
(scaled time in the interval $1$ to $\sim1.7$ in Fig. \ref{fig:time}),
with scaled decay time of $0.23$, and another with a scaled decay
time of $1.02$ when approaching final equilibrium (after scaled time
$\sim2.5$ in Fig. \ref{fig:time}) and an intermediate one in between
(roughly the interval from $1.7$ to $2.5$ in scaled time in Fig.
\ref{fig:time}) as shown in next subsection.

\subsection{Decay of the Condensate}

In Ref. \cite{demidov2008b} is reported by Demidov et al. an analysis
of the decay, towards final equilibrium, of the NEBEC in YIG, after
the external rf-pumping source has been switched off. There are some
differences in the experimental protocol with respect to the one used
in Ref. \cite{demokritov2006}, but we analyse the decay in the conditions
of the latter, to study the dependence (as done in the experiment
of Ref. \cite{demidov2008b}) with the source power.

\begin{figure}[h]
\begin{centering}
\includegraphics[width=0.5\columnwidth]{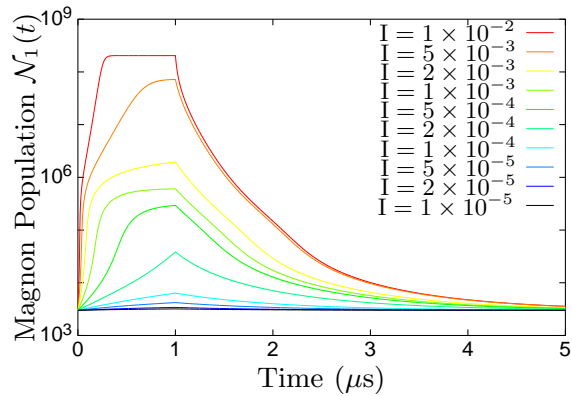}
\par\end{centering}

\caption{\label{fig:decay}(Color online) Time evolution of the condensate ($\mathcal{N}_{1}$)
for different values of the rate of pumping $\mathrm{I}$.}

\end{figure}

Using Eqs. (\ref{eq:N1_evolution}) and (\ref{eq:N2_evolution}) in
Subsection \ref{subsec:BEC-in-YIG}, the one that leads to the results
of Fig. \ref{fig:time}, varying the values of the rate of pumping
$\mathrm{I}$ we obtain the set of curves for the evolution of the
population in the condensate shown in Fig. \ref{fig:decay}.

It may be noticed the interesting result that these curves can be
approximately well adjusted by the law \begin{equation}
\mathcal{N}_{1}(t)-\mathcal{N}_{1}^{(0)}=A\,\exp\left(-\nicefrac{t}{\tau_{\mathrm{A}}}\right)+B\,\exp\left(-\nicefrac{t}{\tau}\right),\label{eq:decay_fit}\end{equation}

The fitting values of the two coefficients, $A$ and $B$, and the
characteristic decay time $\tau_{\mathrm{A}}$ are indicated in Table
1; $\tau$ is the relaxation time to the lattice.

\begin{table}[H]

\caption{\label{tab:decaimento_ajuste}Coefficients $A$, $B$ and $\tau_\mathrm{A}$
obtained from the fitting of Eq. \ref{eq:decay_fit} to the numerical
data presented in Fig. \ref{fig:decay} (it has been taken $\tau=\unit[1]{\mu s}$).}

                                                                                                                                                                                                                                                         \begin{centering}
\begin{tabular}{|c|c|c|c|}
\hline 
$\mathrm{I}$  & $\unit[\tau_{\mathrm{A}}]{(\mu s)}$  & $A$  & $B$ \tabularnewline
\hline
\hline 
$5\times10^{-5}$  & $0,47670$  & $140,10$  & $1064,48$ \tabularnewline
\hline 
$6\times10^{-5}$  & $0,46853$  & $217,96$  & $1316,59$ \tabularnewline
\hline 
$8\times10^{-5}$  & $0,45151$  & $458,56$  & $1869,11$ \tabularnewline
\hline 
$10\times10^{-5}$  & $0,4326$  & $863,2$  & $2496,7$ \tabularnewline
\hline 
$13\times10^{-5}$  & $0,3990$  & $2006,9$  & $3618,7$ \tabularnewline
\hline 
$16\times10^{-5}$  & $0,3568$  & $4460$  & $5048$ \tabularnewline
\hline 
$20\times10^{-5}$  & $0,2849$  & $1,3603\times10^{4}$  & $7786$ \tabularnewline
\hline 
$25\times10^{-5}$  & $0,2131$  & $5,890\times10^{4}$  & $1,278\times10^{4}$ \tabularnewline
\hline 
$32\times10^{-5}$  & $0,21706$  & $1,5904\times10^{5}$  & $1,523\times10^{4}$ \tabularnewline
\hline 
$40\times10^{-5}$  & $0,22634$  & $2,3599\times10^{5}$  & $1,524\times10^{4}$ \tabularnewline
\hline 
$50\times10^{-5}$  & $0,23113$  & $3,0488\times10^{5}$  & $1,517\times10^{4}$ \tabularnewline
\hline 
$63\times10^{-5}$  & $0,23310$  & $3,7937\times10^{5}$  & $1,544\times10^{4}$ \tabularnewline
\hline 
$79\times10^{-5}$  & $0,23263$  & $4,6423\times10^{5}$  & $1,636\times10^{4}$ \tabularnewline
\hline 
$100\times10^{-5}$  & $0,22948$  & $5,765\times10^{5}$  & $1,852\times10^{4}$ \tabularnewline
\hline 
$126\times10^{-5}$  & $0,22314$  & $7,315\times10^{5}$  & $2,29\times10^{4}$ \tabularnewline
\hline 
$158\times10^{-5}$  & $0,2124$  & $9,765\times10^{5}$  & $3,17\times10^{4}$ \tabularnewline
\hline
\end{tabular}
\par\end{centering}

\end{table}

Clearly, we can say that there exist three reasonably well defined
regimes: one immediately after the switching-off of the source, other
at longer decay times when the population is approaching the final
value at equilibrium, and, of course, an intermediate one.

\begin{figure}[h]
\begin{centering}
\includegraphics[width=0.5\columnwidth]{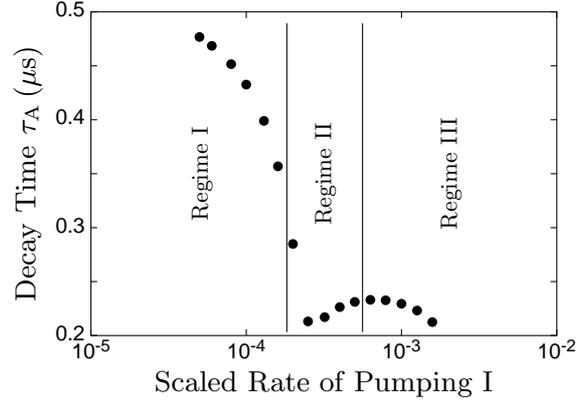}
\par\end{centering}

\caption{\label{fig:decayXintensity}Decay time $\tau_{A}$ of the
condensate as a function of the pumping parameter $\mathrm{I}$.}

\end{figure}

In Fig. \ref{fig:decayXintensity} we indicate the decay time $\tau_\mathrm{A}$
as a function of the scaled rate of pumping $\mathrm{I}$.

It can be noticed a qualitative agreement with the experimental data
of Demidov et al. in figure 3 of reference \cite{demidov2008b}, namely,
a sigmoid-like curve up to $\mathrm{I}\sim1\times10^{-3}$, except
for the central dip in an intermediate region where no experimental
data are reported. We proceed to study in detail the different relaxation
processes that produce these results.

We analyse (using the same notation for rates of the previous section)
the relaxation processes involved: (\emph{i}) the linear relaxation
to the lattice ($\mathrm{L}$-process); (\emph{ii})
the radiative decay ($\mathrm{D}$-process); (\emph{iii}) the conjugated
effect of magnon-magnon interaction and Fr\"{o}hlich effect {[}$(\mathrm{F}+\mathrm{M})$-process{]}.

\begin{figure}[H]
\begin{centering}
\includegraphics[width=0.5\columnwidth]{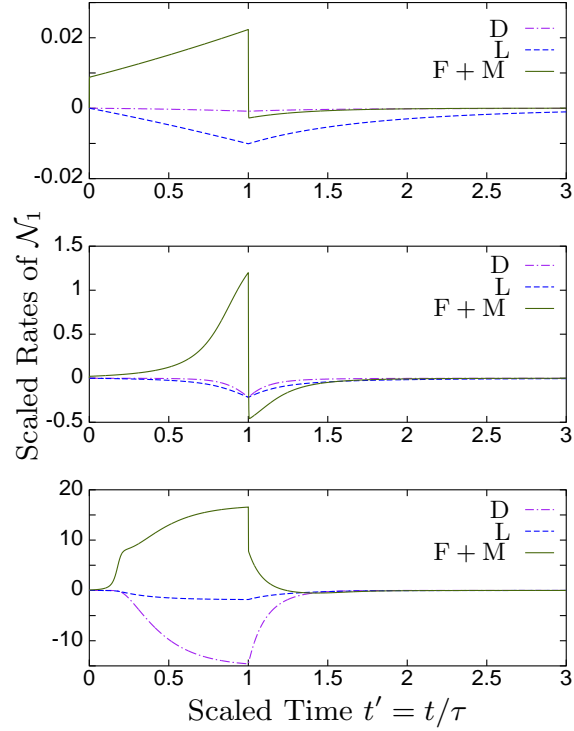}
\par\end{centering}

\caption{\label{fig:decaimentos_taxas} (Color online) Scaled rates of creation and annihilation
of magnons in $R_{1}$ region for different scaled pumping rates.
On the upper figure, with $\mathrm{I}=10^{-4}$, in Regime I, the linear relaxation is
predominant, but the $\mathrm{F}+\mathrm{M}$-process gains relevance.
On the central figure, in Regime II, with $\mathrm{I}=2,5\times10^{-4}$, and the
$\mathrm{F}+\mathrm{M}$-process is the principal relaxation mechanism.
Finally, on the lower figure, we have in Regime III, with $\mathrm{I}=10^{-3}$, where
the $\mathrm{D}$-process dominates the relaxation.}

\end{figure}

At small $\mathrm{I}$, as stated before, the $\mathrm{L}$-process
dominates the decay. But, for intensities such that Eq. \ref{eq:decay_fit}
applies to the populations decay (see figure \ref{fig:decay}), it
can be noticed the increasing relevance of the $\mathrm{F}+\mathrm{M}$-process,
as can be seen in Fig. \ref{fig:decaimentos_taxas}a (Regime I). While
the pumping source is acting, Fr\"{o}hlich and magnon-magnon terms transfer
the energy from the fed magnons ($\mathcal{N}_{2}$) to the ones in
the condensate ($\mathcal{N}_{1}$), but after turning off the source
the flux of energy is inverted, and $\mathrm{F}+\mathrm{M}$-processes
become another mechanism of relaxation of the condensate, thus reducing
$\tau_{A}$. Progressive amplification of the pumping power enlarge
this effect, as shown in Fig. \ref{fig:decayXintensity}. A minimum
$\tau_{A}$ value is achieved for $\mathrm{I}\simeq2,5\times10^{-4}$
(Fig. \ref{fig:decaimentos_taxas}b) and then, for higher $\mathrm{I}$,
the $\mathrm{F}+\mathrm{M}$-process, dominant relaxation process
until that intensity, diminish, and the $\mathrm{D}$-process begins
to increase. Then, the decay time $\tau_{A}$ increases until the
$\mathrm{D}$-process dominates the relaxation (Fig. \ref{fig:decaimentos_taxas}c)
and, after this point, where is initiated the Regime III, $\tau_{A}$
starts to decrease again. For values of $\mathrm{I}$ higher than
$2\times10^{-3}$, the relaxation in the condensate does not follow
Eq. \ref{eq:decay_fit}.

Thus we conclude that, depending on the populations $\mathcal{N}_{1}$
and $\mathcal{N}_{2}$, immediately after turning off the pumping source,
the three distinct contributions indicated rule the condensate relaxation.
Briefly, it can be stated that increasing the pumping source intensity
before turning it off changes the dominant relaxation process in
the following order: \[
\mathrm{L}\to\mathrm{F}+\mathrm{M}\to\mathrm{D},\]
 explaining in this way the unexpected behaviour of $\mathcal{N}_{1}$
while returning to equilibrium.

\section{Concluding Remarks}

In summary, as experimentally evidenced by Demokritov et al. the spin
system in magnetic thin films under excitation by rf-radiation, display
a phenomenon of the type of a Bose-Einstein condensation of the hot
magnons \cite{demokritov2006}. The theoretical analysis performed here, 
in terms of a nonequilibrium
statistical thermo-mechanics, shows that, in fact, this BEC of magnons
follows in the way expected for many-boson systems embedded in a thermal
bath as firstly demonstrated by H. Fr\"{o}hlich \cite{froehlich1970,froehlich1980},
and we may cite him who, in his original work (the case of biopolymers),
stated that 
\begin{quote}
\textquotedblleft{}\ldots{} under appropriate conditions a phenomenon
quite similar to Bose condensation may occur in substances which possess
longitudinal electric modes. If energy is fed into these modes and
thence transferred to other degrees of freedom of the substance then
a stationary state will be reached in which the energy content of
the electric modes is larger than in thermal equilibrium. This excess
energy is found to be channelled into a single mode \textendash{}
exactly as in Bose condensation \textendash{} provided the energy
supply exceeds a critical value. Under these circumstances a random
supply of energy is thus not completely thermalized but partly used
in maintaining a coherent electric wave in the substance.\textquotedblright{}
\end{quote}
Moreover, that long-lived solitons (in the form of a statistically
averaged macroscopic wave function of coherent states) propagate in
the condensate, Finally, it can be noticed that the extended analysis
of the nonequilibrium thermodynamics of the phenomenon described in
the Refs. \cite{fonseca2000} is formally identical to the one that
can be applied to the magnons embedded in the lattice we have considered
here, which is to be extensively described in a forthcoming article.

\ack
Financial support from S\~{a}o Paulo State Research Foundation (FAPESP)
is gratefully acknowledged. RL and AVR are National Research Council
(CNPq) fellows, and FSV is a FAPESP post-doctoral fellow. We thank
Prof. Sergio Rezende (Federal Univ. Pernambuco) for very helpful advice
and for providing us preprints of his recent articles. 

\appendix

\section{The Hamiltonian of the Magnon System}

To introduce the magnon creation (annihilation) operators, we use
the Holstein-Primakoff transformation \cite{keffer1966,akhiezer1968,white1983}:
at first, spin operators are expressed in terms of local bosonic creation
(annihilation) operators, $\hat{a}_{j}^{\dagger}$ ($\hat{a}_{j}$),

\begin{eqnarray}
\hat{S}_{j}^{x}=\frac{\sqrt{2S}}{2}\left[\left(1-\frac{\hat{a}_{j}^{\dagger}\hat{a}_{j}}{2S}\right)^{\nicefrac{1}{2}}\hat{a}_{j}+\hat{a}_{j}^{\dagger}\left(1-\frac{\hat{a}_{j}^{\dagger}\hat{a}_{j}}{2S}\right)^{\nicefrac{1}{2}}\right],\nonumber \\
\hat{S}_{j}^{y}=\frac{\sqrt{2S}}{2i}\left[\left(1-\frac{\hat{a}_{j}^{\dagger}\hat{a}_{j}}{2S}\right)^{\nicefrac{1}{2}}\hat{a}_{j}-\hat{a}_{j}^{\dagger}\left(1-\frac{\hat{a}_{j}^{\dagger}\hat{a}_{j}}{2S}\right)^{\nicefrac{1}{2}}\right],\nonumber \\
\hat{S}_{j}^{z}=S-\hat{a}_{j}^{\dagger}\hat{a}_{j}.\label{eq:transformacao_HP}\end{eqnarray}
 We consider a crystaline material with $N_{\mathrm{c}}$ unit cells,
with the ionic position $\mathbf{R}_{i}$ given by the position $\mathbf{r}_{n}$
of the unit cell $n$ and the internal relative position $\mathbf{d}_{\mu}$,
\begin{equation}
\mathbf{R}_{i}=\mathbf{r}_{n}+\mathbf{d}_{\mu},\end{equation}
 with $\mu$ designating the internal ions ($\mu=1,2,\dots$ until
the number of magnetic ions included in the unit cell), and it is
introduced the Fourier expansion, \begin{equation}
\hat{a}_{j}=\sum_{\mathbf{q},\mu}\frac{\mbox{e}^{i(\mathbf{q}\cdot\mathbf{R}_{j})}}{\sqrt{N_{\mathrm{c}}}}\hat{a}_{\mathbf{q},\mu},\label{eq:fourier}\end{equation}
 being $\mathbf{q}$ the wave vector running over the first Brillouin
zone. Collecting only the quadratic terms of $\ham_{\mathrm{exc}}+\ham_{\mathrm{dip}}+\ham_{\mathrm{Z}}$
{[}Eqs. (\ref{eq:hamiltoniana_de_troca}), (\ref{eq:hamiltoniana_de_dipolo})
and (\ref{eq:hamiltoniana_de_zeeman}){]} in what we call $\ham_{\mathrm{S}}^{(2)}$,
one obtains \begin{eqnarray}
\ham_{\mathrm{S}}^{(2)}& = \sum_{\mathbf{q},\mu,\mu'}\Biggl\{\mathrm{A}'_{\mu\mu'}(\mathbf{q})\,\hat{a}_{\mathbf{q},\mu}^{\dagger}\hat{a}_{\mathbf{q},\mu'}+\nonumber \\
 &  +\frac{\mathrm{B}_{\mu\mu'}'^{*}(\mathbf{q})}{2}\hat{a}_{\mathbf{q},\mu}^{\dagger}\hat{a}_{-\mathbf{q},\mu'}^{\dagger}+\frac{\mathrm{B}'_{\mu\mu'}(\mathbf{q})}{2}\hat{a}_{\mathbf{q},\mu}\hat{a}_{-\mathbf{q},\mu'}\Biggr\},\label{eq:eq:ham_quad_pre_bogoliubov}\end{eqnarray}
with \begin{eqnarray}
\mathrm{A}'_{\mu\mu'}(\mathbf{q})& = \sum_{n,n'}\left\{ \mathrm{A}_{1}(\mathbf{r}_{nn'}+\mathbf{d}_{\mu\mu'})\mbox{e}^{i\left[\mathbf{q}\cdot(\mathbf{r}_{nn'}+\mathbf{d}_{\mu\mu'})\right]}+\right.\nonumber \\
 & \qquad\left.+\mathrm{A}_{2}(\mathbf{r}_{nn'}+\mathbf{d}_{\mu\mu'})\right\} ,\label{eq:somas_chatas_A}\end{eqnarray}
 \begin{eqnarray}
\mathrm{B}'_{\mu\mu'}(\mathbf{q})=\sum_{n,n'}\mathrm{B}(\mathbf{r}_{nn'}+\mathbf{d}_{\mu\mu'})\mbox{e}^{i\left[\mathbf{q}\cdot(\mathbf{r}_{nn'}+\mathbf{d}_{\mu\mu'})\right]},\label{eq:somas_chatas_B}\end{eqnarray}
 where $\mathbf{r}_{nn'}=\mathbf{r}_{n'}-\mathbf{r}_{n}$ and with
$(n,\mu)\neq(n',\mu')$, and \begin{eqnarray}
\mathrm{A}_{1}(\mathbf{R}_{ij})=-2SJ(R_{ij})-\frac{(g\mu_{\mathrm{B}})^{2}S}{2R_{ij}^{5}}[2R_{ij}^{2}-3(R_{ij}^{z})^{2}],\label{eq:coeficiente_A1}\\
\mathrm{A}_{2}(\mathbf{R}_{ij})=2SJ(R_{ij})+\frac{g\mu_{\mathrm{B}}\mathrm{H}_{0}}{N}-\frac{(g\mu_{\mathrm{B}})^{2}S}{R_{ij}^{5}}[R_{ij}^{2}-3(R_{ij}^{z})^{2}],\label{eq:coeficiente_A2}\\
\mathrm{B}(\mathbf{R}_{ij})=-\frac{3(g\mu_{\mathrm{B}})^{2}S}{2}\frac{(R_{ij}^{-})^{2}}{R_{ij}^{5}}.\label{eq:coeficiente_B}\end{eqnarray}

The diagonalization of $\ham_{\mathrm{S}}^{(2)}$ of Eq. \ref{eq:eq:ham_quad_pre_bogoliubov}
is done through the introduction of the magnon creation and annihilation
operators $\hat{c}_{\mathbf{q},\gamma}^{\dagger}$ and $\hat{c}_{\mathbf{q},\gamma}$,
as linear combinations of $\hat{a}_{\mathbf{q},\mu}^{\dagger}$ and
$\hat{a}_{\mathbf{q},\mu}$ in such a way that \cite{keffer1966,harris1963,cherepanov1993}
\begin{equation}
\ham_{\mathrm{S}}^{(2)}=\sum_{\mathbf{q},\gamma}\hbar\omega_{\mathbf{q},\gamma}\hat{c}_{\mathbf{q},\gamma}^{\dagger}\hat{c}_{\mathbf{q},\gamma}.\label{eq:hamiltoniana_magnons_ramos}\end{equation}

The resulting magnons, with energy $\hbar\omega_{\mathbf{q},\gamma}$
and group velocity $\nabla_{\mathbf{q}}\omega_{\mathbf{q},\gamma}$,
are grouped in branches indicated by $\gamma$, in the case of more
than one magnetic ion per unitary cell (see Refs. \cite{harris1963,cherepanov1993}
for the YIG). In the experiments with thin films of YIG analyzed here,
only low frequency magnons are excited ($\lesssim\unit[10]{GHz}$),
justifying the omission of all the branches but the acoustic one.
In this sense, only one effective spin per unit cell is considered,
$\mu$ and $\gamma$ are omitted and we write for $\ham_{\mathrm{S}}^{(2)}$
of Eq. (\ref{eq:hamiltoniana_magnons_ramos}) \begin{equation}
\ham_{\mathrm{S}}^{(2)}\simeq\sum_{\mathbf{q}}\hbar\omega_{\mathbf{q}}\hat{c}_{\mathbf{q}}^{\dagger}\hat{c}_{\mathbf{q}}.\label{eq:hamiltoniana_magnonslivres}\end{equation}

In this situation of taking into account only the acoustic magnons,
Eq. (\ref{eq:hamiltoniana_magnonslivres}) follows from Eq. (\ref{eq:eq:ham_quad_pre_bogoliubov}),
when in the latter we take one effective spin per unit cell, after
using the so-called Bogoliubov transformation, \cite{keffer1966,akhiezer1968},
\begin{eqnarray}
a_{\mathbf{q}} & = & u_{\mathbf{q}}c_{\mathbf{q}}+v_{\mathbf{q}}^{*}c_{-\mathbf{q}}^{\dagger},\nonumber \\
a_{-\mathbf{q}}^{\dagger} & = & u_{-\mathbf{q}}^{*}c_{-\mathbf{q}}^{\dagger}+v_{-\mathbf{q}}c_{\mathbf{q}},\label{eq:transformacao_bogoliubov}\end{eqnarray}
 with $u_{\mathbf{q}}$ and $v_{\mathbf{q}}$ functions of $\mathbf{q}$.
It can be shown that \begin{eqnarray}
u_{\mathbf{q}}=\sqrt{\frac{\mathrm{A}'(\mathbf{q})+\hbar\omega_{\mathbf{q}}}{2\hbar\omega_{\mathbf{q}}}}, \quad v_{\mathbf{q}}=\frac{\mathrm{B}'(\mathbf{q})}{\left|\mathrm{B}'(\mathbf{q})\right|}\sqrt{\frac{\mathrm{A}'(\mathbf{q})-\hbar\omega_{\mathbf{q}}}{2\hbar\omega_{\mathbf{q}}}},\label{eq:coeficientes_bogoliubov}\end{eqnarray}
 and the (acoustic) magnon dispersion relation is given by \begin{equation}
\hbar\omega_{\mathbf{q}}=\sqrt{\left[\mathrm{A}'(\mathbf{q})\right]^{2}-\left|\mathrm{B}'(\mathbf{q})\right|^{2}},\label{eq:dispersao}\end{equation}
 and Refs. \cite{rezende2009,kreisel2009} presents recent studies
on this dispersion relation in cases of thin films of YIG.

The Holstein-Primakoff and Bogoliubov transformations are then applied
to the non-quadratic terms of $\ham_{\mathrm{exc}}+\ham_{\mathrm{dip}}+\ham_{\mathrm{Z}}$,
and the magnon-magnon interaction term is then obtained. Retaining
only the fourth order scattering terms, the magnon-magnon interaction
which contributes to the Hamiltonian is given by \[
\ham_{\mathrm{MM}}=\sum_{\mathbf{q},\mathbf{q}_{1},\mathbf{q}_{2}}\mathcal{V}_{\mathbf{q},\mathbf{q}_{1},\mathbf{q}_{2}}\hat{c}_{\mathbf{q}}^{\dagger}\hat{c}_{\mathbf{q}_{1}}^{\dagger}\hat{c}_{\mathbf{q}_{2}}\hat{c}_{\mathbf{q}+\mathbf{q}_{1}-\mathbf{q}_{2}}.\]

Phonon and photon creation and annihilation operators are introduced
in similar ways. The Hamiltonian of the free phonons is \begin{equation}
\ham_{\mathrm{L}}=\sum_{\mathbf{k}}\hbar\Omega_{\mathbf{k}}\left(\hat{b}_{\mathbf{k}}^{\dagger}\hat{b}_{\mathbf{k}}+\frac{1}{2}\right),\label{eq:hamiltoniano_fonons}\end{equation}
 with $\Omega_{\mathbf{k}}$ being their dispersion relation (we recall
that only acoustic phonons were considered; polarization is implicit).
The phonon creation and annihilation operators, $\hat{b}_{\mathbf{k}}^{\dagger}$
and $\hat{b}_{\mathbf{k}}$, are related to the displacement of the
effective magnetic ion around the equilibrium position $\mathbf{r}_{n}$,
which is \begin{equation}
\mathbf{x}_{n}=\left(\frac{\hbar}{2N_{\mathrm{c}}}\right)^{\frac{1}{2}}\sum_{\mathbf{k}}\frac{\mathbf{e}(\mathbf{k})}{\sqrt{\Omega_{\mathbf{k}}}}\left(\hat{b}_{\mathbf{k}}\mbox{e}^{i\mathbf{k}\cdot\mathbf{r}_{n}}+\hat{b}_{\mathbf{k}}^{\dagger}\mbox{e}^{-i\mathbf{k}\cdot\mathbf{r}_{n}}\right),\label{eq:displacement_phonons}\end{equation}
 and $\mathbf{e}(\mathbf{k})$ is the polarization versor (cf. Ref.
\cite{akhiezer1968}). Using Eq. \ref{eq:displacement_phonons} in
Eq. \ref{eq:hamiltoniana_spinrede}, we obtain the for magnon-phonon
interaction 

\begin{eqnarray}
\ham_{\mathrm{SL}}& = {\displaystyle \sum_{\mathbf{q},\mathbf{k}\neq0}}(\hat{b}_{\mathbf{k}}+\hat{b}_{-\mathbf{k}}^{\dagger})\Big\{ \mathcal{F}_{\mathbf{q},\mathbf{k}}\hat{c}_{\mathbf{q}}^{\dagger}\hat{c}_{\mathbf{q}-\mathbf{k}}+\mathcal{L}_{\mathbf{q},\mathbf{k}}\hat{c}_{\mathbf{q}}^{\dagger}\hat{c}_{\mathbf{k}-\mathbf{q}}^{\dagger}+\nonumber \\
 & \qquad\qquad\qquad\qquad+\mathcal{L}_{\mathbf{q},-\mathbf{k}}^{*}\hat{c}_{\mathbf{q}}\hat{c}_{-\mathbf{k}-\mathbf{q}}\Big\} +\nonumber \\
 & +{\displaystyle \sum_{\mathbf{q},\mathbf{k}\neq0}}\Big\{ \mathcal{R}_{\mathbf{q},\mathbf{k}}\hat{b}_{\mathbf{k}}^{\dagger}\hat{b}_{\mathbf{k}-\mathbf{q}}+\mathcal{R}_{\mathbf{q},\mathbf{k}}^{+}\hat{b}_{\mathbf{k}}^{\dagger}\hat{b}_{\mathbf{q}-\mathbf{k}}^{\dagger}+\nonumber \\
 & \qquad+\mathcal{R}_{-\mathbf{q},-\mathbf{k}}^{+*}\hat{b}_{-\mathbf{k}}\hat{b}_{\mathbf{k}-\mathbf{q}}\Big\} (\hat{c}_{\mathbf{q}}+\hat{c}_{-\mathbf{q}}^{\dagger}),\end{eqnarray}
 where $\mathcal{F}_{\mathbf{q},\mathbf{k}}$, $\mathcal{L}_{\mathbf{q},\mathbf{k}}$
and $\mathcal{R}_{\mathbf{q},\mathbf{k}}^{(\pm)}$ are the resulting
coefficients (representing the interaction coupling intensities).
We write for the field generated by photons of the electromagnetic
fields (from the source and black-body radiation) \cite{landaulifshitz_QE}
\begin{equation}
\mathbf{H}(\mathbf{r})=\sum_{\alpha,\mathbf{p}}(\mathbf{H}_{\alpha,\mathbf{p}}(\mathbf{r})\,\hat{d}_{\alpha,\mathbf{p}}+\mathbf{H}_{\alpha,\mathbf{p}}^{*}(\mathbf{r})\,\hat{d}_{\alpha,\mathbf{p}}^{\dagger}),\label{eq:quantizacao_fotons}\end{equation}
 where $\hat{d}_{\alpha,\mathbf{q}}$ ($\hat{d}_{\alpha,\mathbf{q}}^{\dagger}$)
are the photon creation (annihilation) operators, \begin{equation}
\mathbf{H}_{\alpha,\mathbf{p}}(\mathbf{r})=i\mathbf{p}\times\mathbf{A}_{\alpha,\mathbf{p}}(\mathbf{r}),\qquad\mathbf{A}_{\alpha,\mathbf{p}}(\mathbf{r})=\sqrt{\frac{2\pi}{\zeta_{\mathbf{p}}}}\mbox{e}^{i\mathbf{p}\cdot\mathbf{r}}\mathbf{e}^{(\alpha)},\end{equation}
 $\zeta_{\mathbf{p}}$ is the photon angular frequency, $\mathbf{p}$
its linear moment and $\mathbf{e}^{(\alpha)}$ the polarization vector
(the different polarizations are indexed by $\alpha$).

After some algebra, we obtain for the magnon-photon interaction of
Eq. \ref{eq:hamiltoniana_spinluz} the expression \begin{eqnarray}
\ham_{\mathrm{SR}}& = \sum_{\alpha,\mathbf{p}}(\hat{d}_{\alpha,\mathbf{p}}+\hat{d}_{\alpha,-\mathbf{p}}^{\dagger})\left(\mathcal{S}_{\alpha,\mathbf{p}}^{\perp*}\hat{c}_{\mathbf{p}}^{\dagger}+\mathcal{S}_{\alpha,-\mathbf{p}}^{\perp}\hat{c}_{-\mathbf{p}}\right)+\nonumber \\
 & +\sum_{\alpha,\mathbf{p},\mathbf{q}}(\hat{d}_{\alpha,\mathbf{p}}+\hat{d}_{\alpha,-\mathbf{p}}^{\dagger})\Big\{ \mathcal{S}_{\alpha,\mathbf{q},\mathbf{p}}^{\parallel\mathrm{a}}\hat{c}_{\mathbf{q}}^{\dagger}\hat{c}_{\mathbf{q}-\mathbf{p}}+v_{\mathbf{q}}v_{\mathbf{q}-\mathbf{p}}^{*}+\nonumber \\
 & \qquad+\mathcal{S}_{\alpha,\mathbf{q},\mathbf{p}}^{\parallel\mathrm{b}}\hat{c}_{\mathbf{q}}^{\dagger}\hat{c}_{\mathbf{p}-\mathbf{q}}^{\dagger}+\mathcal{S}_{\alpha,-\mathbf{q},-\mathbf{p}}^{\parallel\mathrm{b}*}\hat{c}_{-\mathbf{q}}\hat{c}_{\mathbf{q}-\mathbf{p}}\Big\} ,\end{eqnarray}
 where $\mathcal{S}_{\alpha,\mathbf{p}}^{\perp*}$, $\mathcal{S}_{\alpha,\mathbf{q},\mathbf{p}}^{\parallel\mathrm{a}}$
and $\mathcal{S}_{\alpha,\mathbf{q},\mathbf{p}}^{\parallel\mathrm{b}}$
are intensity coupling coefficients. Considering that the external
pumping source and black-body radiation are the origins of the electromagnetic
fields, we can distinguish respectively their operators as $\hat{d}_{\alpha,\mathbf{q}}^{\mathrm{S}}$
and $\hat{d}_{\alpha,\mathbf{q}}^{\,\mathrm{T}}$, and the energy
of the black-body radiation is \begin{equation}
\ham_{\mathrm{R}}=\sum_{\alpha,\mathbf{p}}\hbar\zeta_{\mathbf{p}}\left(\hat{d}_{\alpha,\mathbf{q}}^{\,\mathrm{T}\dagger}\hat{d}_{\alpha,\mathbf{q}}^{\,\mathrm{T}}+\frac{1}{2}\right).\label{eq:hamiltoniano_fotons}\end{equation}

The complete Hamiltonian is then written, in second quantization form,
as

\begin{eqnarray}
\fl\ham=&\sum_{\mathbf{q}}\hbar\omega_{\mathbf{q}}\hat{c}_{\mathbf{q}}^{\dagger}\hat{c}_{\mathbf{q}}+\sum_{\mathbf{k}}\hbar\Omega_{\mathbf{k}}\hat{b}_{\mathbf{k}}^{\dagger}\hat{b}_{\mathbf{k}}+\sum_{\alpha,\mathbf{p}}\hbar\zeta_{\mathbf{p}}\hat{d}_{\alpha,\mathbf{q}}^{\dagger}\hat{d}_{\alpha,\mathbf{q}}+\nonumber \\
\fl &+\sum_{\mathbf{q},\mathbf{q}_{1},\mathbf{q}_{2}}\mathcal{V}_{\mathbf{q},\mathbf{q}_{1},\mathbf{q}_{2}}\hat{c}_{\mathbf{q}}^{\dagger}\hat{c}_{\mathbf{q}_{1}}^{\dagger}\hat{c}_{\mathbf{q}_{2}}\hat{c}_{\mathbf{q}+\mathbf{q}_{1}-\mathbf{q}_{2}}+\nonumber \\
\fl &+{\displaystyle \sum_{\mathbf{q},\mathbf{k}\neq0}}(\hat{b}_{\mathbf{k}}+\hat{b}_{-\mathbf{k}}^{\dagger})\left\{ \mathcal{F}_{\mathbf{q},\mathbf{k}}\hat{c}_{\mathbf{q}}^{\dagger}\hat{c}_{\mathbf{q}-\mathbf{k}}+\mathcal{L}_{\mathbf{q},\mathbf{k}}\hat{c}_{\mathbf{q}}^{\dagger}\hat{c}_{\mathbf{k}-\mathbf{q}}^{\dagger}+\mathcal{L}_{\mathbf{q},-\mathbf{k}}^{*}\hat{c}_{\mathbf{q}}\hat{c}_{-\mathbf{k}-\mathbf{q}}\right\} +\nonumber \\
\fl &+{\displaystyle \sum_{\mathbf{q},\mathbf{k}\neq0}}\left\{ \mathcal{R}_{\mathbf{q},\mathbf{k}}\hat{b}_{\mathbf{k}}^{\dagger}\hat{b}_{\mathbf{k}-\mathbf{q}}+\mathcal{R}_{\mathbf{q},\mathbf{k}}^{+}\hat{b}_{\mathbf{k}}^{\dagger}\hat{b}_{\mathbf{q}-\mathbf{k}}^{\dagger}+\mathcal{R}_{-\mathbf{q},-\mathbf{k}}^{+*}\hat{b}_{-\mathbf{k}}\hat{b}_{\mathbf{k}-\mathbf{q}}\right\} (\hat{c}_{\mathbf{q}}+\hat{c}_{-\mathbf{q}}^{\dagger})+\nonumber \\
\fl &+\sum_{\alpha,\mathbf{p}}(\hat{d}_{\alpha,\mathbf{p}}+\hat{d}_{\alpha,-\mathbf{p}}^{\dagger})\left(\mathcal{S}_{\alpha,\mathbf{p}}^{\perp*}\hat{c}_{\mathbf{p}}^{\dagger}+\mathcal{S}_{\alpha,-\mathbf{p}}^{\perp}\hat{c}_{-\mathbf{p}}\right)+\nonumber \\
\fl &+\sum_{\alpha,\mathbf{p},\mathbf{q}}(\hat{d}_{\alpha,\mathbf{p}}+\hat{d}_{\alpha,-\mathbf{p}}^{\dagger})\left\{ \mathcal{S}_{\alpha,\mathbf{q},\mathbf{p}}^{\parallel\mathrm{a}}\hat{c}_{\mathbf{q}}^{\dagger}\hat{c}_{\mathbf{q}-\mathbf{p}}+\mathcal{S}_{\alpha,\mathbf{q},\mathbf{p}}^{\parallel\mathrm{b}}\hat{c}_{\mathbf{q}}^{\dagger}\hat{c}_{\mathbf{p}-\mathbf{q}}^{\dagger}+\mathcal{S}_{\alpha,\mathbf{q},-\mathbf{p}}^{\parallel\mathrm{b}*}\hat{c}_{-\mathbf{q}}\hat{c}_{\mathbf{q}-\mathbf{p}}\right\}.\nonumber \\
\fl&\label{eq:ham_linha_eq_cineticas}
\end{eqnarray}

\section{The Nonequilibrium Statistical Operator}

The nonequilibrium statistical operator is given by \cite{luzzilivro2002,luzzi2006,zubarev1996,kuzensky2009,akhiezer1981,mclennan1963}
\begin{equation}
\hat{\mathscr R}_{\varepsilon}(t)=\hat{\varrho}_{\varepsilon}(t)\times\hat{\varrho}_{\mathrm{B}},\label{eq:operador_completo}\end{equation}
where \begin{equation}
\hat{\varrho}_{\varepsilon}(t)=\exp\left\{ \ln\hat{\bar{\rho}}(t,0)-\int_{t_{0}}^{t}dt'\mbox{e}^{\varepsilon\left(t'-t\right)}\frac{d}{dt'}\ln\hat{\bar{\rho}}(t',t'-t)\right\} \label{eq:rho_final}\end{equation}
 is the nonequilibrium statistical operator of the magnon system,
with the auxiliary statistical operator $\hat{\bar{\rho}}$ (also
called {}``instantaneous quasi-equilibrium operator''), depending
(superoperator) on the basic microvariables of set \ref{eq:microdynamical_variables_3},
given by \begin{eqnarray}
\hat{\bar{\rho}}(t,0)=\; & \exp\Biggl\{-\Phi(t)-{\displaystyle \sum_{\mathbf{q}}}\left[F_{\mathbf{q}}(t)\hat{c}_{\mathbf{q}}^{\dagger}\hat{c}_{\mathbf{q}}+\right.\nonumber\\
 & \qquad+\phi_{\mathbf{q}}(t)\hat{c}_{\mathbf{q}}+\phi_{\mathbf{q}}^{*}(t)\hat{c}_{\mathbf{q}}^{\dagger}+\nonumber\\
&\left.\qquad+\varphi_{\mathbf{q}}(t)\hat{c}_{\mathbf{q}}\hat{c}_{-\mathbf{q}}+\varphi_{\mathbf{q}}^{*}(t)\hat{c}_{\mathbf{q}}^{\dagger}\hat{c}_{-\mathbf{q}}^{\dagger}\right]\Biggr\} \end{eqnarray}
 after recalling that we haze neglected local inhomogeneities. In
$\bar{\rho}(t',t'')$the first term in the argument, $t'$, refers
to the evolution in time of the nonequilibrium thermodynamic state
of the system (i.e., of the nonequilibrium thermodynamic variables
$F_{\mathbf{q}}(t)$, $\phi_{\mathbf{q}}(t)$ and $\varphi_{\mathbf{q}}(t)$),
and the second, $t''$, to the evolution of the microdynamical variables
$\hat{\mathcal{N}}_{\mathbf{q}}$, $\hat{c}_{\mathbf{q}}$ and $\hat{\sigma}_{\mathbf{q}}$
in Heisenberg representation. $\varepsilon$ is a positive infinitesimal
that goes to $+0$ after calculation of average values has been performed.
Moreover, $\Phi(t)$ ensures the normalization of the probability
distributions, and plays the role of the logarithm of a nonequilibrium
partition function: $\Phi(t)=\ln\bar{Z}(t)$. It is verified that
\begin{equation}
\mathcal{N}_{\mathbf{q}}(t)=\frac{\delta\ln\bar{Z}}{\delta F_{\mathbf{q}}(t)},\end{equation}
 and similarly for the other quantities, in complete analogy with
the situation in equilibrium, meaning that $F_{\mathbf{q}}(t)$ and
the others in set (\ref{eq:thermovariables}) are the nonequilibrium
thermodynamic variables said conjugated to the basic variables in
set (\ref{eq:macrovariables}). Furthermore, it is verified that \begin{equation}
\hat{\varrho}_{\varepsilon}(t)=\hat{\bar{\rho}}(t,0)+\hat{\varrho}'_{\varepsilon}(t),\label{eq:rho_decomposto}\end{equation}
 where $\hat{\varrho}'_{\varepsilon}(t)$ incorporates irreversibility
and historicity. In Eq. \ref{eq:operador_completo} \begin{equation}
\hat{\varrho}_{\mathrm{B}}=Z_{\mathrm{B}}^{-1}(T_{0})\exp\left\{ \frac{\ham_{\mathrm{L}}+\ham_{\mathrm{R}}}{k_{\mathrm{B}}T_{0}}\right\} \label{eq:rho_canonico}\end{equation}
 is the canonical distribution function of the phonons and photons
in equilibrium at temperature $T_{0}$.

\section{The Evolution Equations for the Basic Variables}

The macrovariable $Q_{j}$ {[}a generic expression for those of the
set indicated in Eq. (\ref{eq:macrovariables}){]}, related to the
microdynamical one $\hat{P_{j}}$ {[}of the set in Eq. (\ref{eq:microdynamical_variables_3}){]},
has its evolution given by the Heisenberg equation of motion weighted
with the nonequilibrium statistical operator $\hat{\mathscr R}_{\varepsilon}(t)$,
namely \begin{equation}
\frac{\partial}{\partial t}Q_{j}(t)=\frac{1}{i\hbar}\mbox{Tr}\left\{ \left[\hat{P}_{j},\ham\right]\,\hat{\mathscr R}_{\varepsilon}(t)\right\} ,\label{eq:q_evolution}\end{equation}
 which, considering Eq. (\ref{eq:operador_completo}) and that, according
to \ref{eq:rho_decomposto}, $\hat{\varrho}'_{\varepsilon}(t)=\hat{\varrho}_{\varepsilon}(t)-\hat{\bar{\varrho}}(t,0)$,
can be rewritten as \begin{equation}
\frac{\partial}{\partial t}Q_{j}(t)=J_{Q_{j}}^{(0)}(t)+J_{Q_{j}}^{(1)}(t)+\mathcal{J}_{Q_{j}}(t),\label{eq:q_evolution_Js}\end{equation}
with\begin{equation}
J_{Q_{j}}^{(0)}(t)\equiv\frac{1}{i\hbar}\mbox{Tr}\left\{ \left[\hat{P}_{j},\ham^{0}\right]\,\hat{\bar{\varrho}}(t,0)\times\varrho_{\mathrm{B}}\right\} ,\end{equation}
\begin{equation}
J_{Q_{j}}^{(1)}(t)\equiv\frac{1}{i\hbar}\mbox{Tr}\left\{ \left[\hat{P}_{j},\ham'\right]\,\hat{\bar{\varrho}}(t,0)\times\varrho_{\mathrm{B}}\right\} \end{equation}
and \begin{equation}
\mathcal{J}_{Q_{j}}(t)=\frac{1}{i\hbar}\mbox{Tr}\left\{ \left[\hat{P}_{j},\ham'\right]\,\hat{\varrho}'_{\varepsilon}(t,0)\times\varrho_{\mathrm{B}}\right\} .\end{equation}
 In the Markovian approximation we have that \begin{eqnarray}
\!\mathcal{J}_{Q_{j}}(t)\simeq \: & J_{Q_{j}}^{(2)}(t)\equiv \nonumber \\
 & \frac{1}{(i\hbar)^{2}}\int_{-\infty}^{t}dt'\mbox{ e}^{\varepsilon(t'-t)}\times\nonumber \\
 & \times\mbox{ Tr}\left\{ \left[\ham'(t'-t)_{0},[\ham',\hat{P}_{j}]\right]\,\hat{\bar{\varrho}}(t,0)\times\varrho_{\mathrm{B}}\right\} +\nonumber \\
 & +\frac{1}{i\hbar}\sum_{\ell}\int_{-\infty}^{t}dt'\mbox{ e}^{\varepsilon(t'-t)}\times\nonumber \\
 & \times\mbox{Tr}\left\{ [\ham'(t'-t)_{0},\hat{P}_{j}]\,\hat{\bar{\varrho}}(t,0)\times\varrho_{\mathrm{B}}\right\} \frac{\delta J_{Q_{j}}^{(1)}(t)}{\delta Q_{\ell}(t)},\end{eqnarray} 
 and the evolution of $Q_{j}$, is thus expressed only in terms of
average values weighted with $\hat{\bar{\varrho}}(t,0)$, where $\delta$
stands for functional derivative.

The evolution equations for the amplitudes are \begin{equation}
\frac{\partial}{\partial t}\left\langle \hat{c}_{\mathbf{q}}|t\right\rangle =J_{c_{\mathbf{q}}}^{(0)}(t)+J_{c_{\mathbf{q}}}^{(1)}(t)+J_{c_{\mathbf{q}}}^{(2)}(t),\label{eq:q_evolution_Js-1}\end{equation}
 with \begin{equation}
J_{c_{\mathbf{q}}}^{(0)}(t)=-i\omega_{\mathbf{q}}\left\langle \hat{c}_{\mathbf{q}}|t\right\rangle \label{eq:J_0_c_q_apendice}\end{equation}
 (the precession term in Mori's terminology \cite{mori1965}), 

\begin{eqnarray}
J_{c_{\mathbf{q}}}^{(1)}(t)& = \frac{2}{i\hbar}\sum_{\mathbf{q}_{1},\mathbf{q}_{2}}\mathcal{V}_{\mathbf{q},\mathbf{q}_{1},\mathbf{q}_{2}}\left\langle \hat{c}_{\mathbf{q}_{1}}^{\dagger}\hat{c}_{\mathbf{q}_{2}}\hat{c}_{\mathbf{q}+\mathbf{q}_{1}-\mathbf{q}_{2}}|t\right\rangle =\nonumber \\
& = \frac{2}{i\hbar}\sum_{\mathbf{q}_{1},\mathbf{q}_{2}}\mathcal{V}_{\mathbf{q},\mathbf{q}_{1},\mathbf{q}_{2}}\left\langle \hat{c}_{\mathbf{q}_{1}}^{\dagger}|t\right\rangle \left\langle \hat{c}_{\mathbf{q}_{2}}|t\right\rangle \left\langle \hat{c}_{\mathbf{q}+\mathbf{q}_{1}-\mathbf{q}_{2}}|t\right\rangle +\nonumber \\
 & +\frac{4}{i\hbar}\sum_{\mathbf{q}_{1}}\mathcal{V}_{\mathbf{q},\mathbf{q}_{1},\mathbf{q}_{1}}\left[\mathcal{N}_{\mathbf{q}_{1}}(t)-\left|\left\langle \hat{c}_{\mathbf{q}_{1}}|t\right\rangle \right|^{2}\right]\left\langle \hat{c}_{\mathbf{q}}|t\right\rangle +\nonumber \\
 & +\frac{4}{i\hbar}\sum_{\mathbf{q}_{1}}\mathcal{V}_{\mathbf{q},-\mathbf{q},\mathbf{q}_{1}}\left[\sigma_{\mathbf{q}_{1}}(t)-\left\langle \hat{c}_{\mathbf{q}_{1}}|t\right\rangle \left\langle \hat{c}_{-\mathbf{q}_{1}}|t\right\rangle \right]\left\langle \hat{c}_{-\mathbf{q}}^{\dagger}|t\right\rangle ,\label{eq:Jc_1}\end{eqnarray}
 arising out of the magnon-magnon interaction, and \[
J_{c_{\mathbf{q}}}^{(2)}(t)=J_{c_{\mathbf{q}}}^{(2)}(t)_{\mathrm{I}}+J_{c_{\mathbf{q}}}^{(2)}(t)_{\mathrm{II}},\]
 where \begin{equation}
J_{c_{\mathbf{q}}}^{(2)}(t)_{\mathrm{I}}=J_{c_{\mathbf{q}}}^{(2)}(t)_{\mathrm{I}}^{\mathrm{MM}}+J_{c_{\mathbf{q}}}^{(2)}(t)_{\mathrm{I}}^{\mathrm{SL}}+J_{c_{\mathbf{q}}}^{(2)}(t)_{\mathrm{I}}^{\mathrm{SR}},\label{eq:J2_c_I}\end{equation}

\begin{eqnarray}
J_{c_{\mathbf{q}}}^{(2)}(t)_{\mathrm{II}}& = \frac{1}{i\hbar}\sum_{\ell}\int_{-\infty}^{t}dt'\mbox{ e}^{\varepsilon(t'-t)}\times\nonumber \\
 & \hspace{-2em}\times\mbox{Tr}\left\{ \left[\ham'(t'-t)_{0},\hat{P}_{\ell}\right]\,\hat{\bar{\varrho}}(t,0)\times\varrho_{\mathrm{B}}\right\} \frac{\delta J_{c_{\mathbf{q}}}^{(1)}(t)}{\delta Q_{\ell}(t)},\label{eq:J2_c_II}\end{eqnarray}
with\begin{eqnarray}
J_{c_{\mathbf{q}}}^{(2)}(t)_{\mathrm{I}}^{\mathrm{MM}}& = \frac{1}{(i\hbar)^{2}}\int_{-\infty}^{t}dt'\mbox{ e}^{\varepsilon(t'-t)}\times\nonumber \\
 & \hspace{-2em}\times\mbox{ Tr}\left\{ \left[\ham_{\mathrm{MM}}(t'-t)_{0},[\ham_{\mathrm{MM}},\hat{c}_{\mathbf{q}}]\right]\,\hat{\bar{\varrho}}(t,0)\times\varrho_{\mathrm{B}}\right\} ,\label{eq:J2_c_MM}\end{eqnarray}
\begin{eqnarray}
J_{c_{\mathbf{q}}}^{(2)}(t)_{\mathrm{I}}^{\mathrm{SL}}& = \frac{1}{(i\hbar)^{2}}\int_{-\infty}^{t}dt'\mbox{ e}^{\varepsilon(t'-t)}\times\nonumber \\
 & \hspace{-2em}\times\mbox{ Tr}\left\{ \left[\ham_{\mathrm{SL}}(t'-t)_{0},[\ham_{\mathrm{SL}},\hat{c}_{\mathbf{q}}]\right]\,\hat{\bar{\varrho}}(t,0)\times\varrho_{\mathrm{B}}\right\} \label{eq:J2_c_SL}\end{eqnarray}
and \begin{eqnarray}
J_{c_{\mathbf{q}}}^{(2)}(t)_{\mathrm{I}}^{\mathrm{SR}}& = \frac{1}{(i\hbar)^{2}}\int_{-\infty}^{t}dt'\mbox{ e}^{\varepsilon(t'-t)}\times\nonumber \\
 & \hspace{-2em}\times\mbox{ Tr}\left\{ \left[\ham_{\mathrm{SR}}(t'-t)_{0},[\ham_{\mathrm{SR}},\hat{c}_{\mathbf{q}}]\right]\,\hat{\bar{\varrho}}(t,0)\times\varrho_{\mathrm{B}}\right\} .\label{eq:J2_c_SR}\end{eqnarray}

The scattering integrals $J_{c_{\mathbf{q}}}^{(2)}(t)_{\mathrm{I}}$
and $J_{c_{\mathbf{q}}}^{(2)}(t)_{\mathrm{II}}$ should be written
in terms of populations, amplitudes and pairs of magnons, and in which
double commutator of Eqs. (\ref{eq:J2_c_MM}) - (\ref{eq:J2_c_SR})
and the $\ell$-sum of Eq. (\ref{eq:J2_c_II}) generate a huge number
of terms, of which we analyze a particular one for illustration, say

\begin{eqnarray}
\fl J_{c_{\mathbf{q}}}^{(2)}(t)_{\mathrm{I}}^{\mathrm{MM}}=&-8\hbar^{-2}&\sum_{\mathbf{q}_{1},\mathbf{q}_{2},\mathbf{q}_{3},\mathbf{q}_{4}}\mathcal{V}_{\mathbf{q},\mathbf{q}_{1},\mathbf{q}_{2}}\mathcal{V}_{\mathbf{q}_{3},\mathbf{q}_{2},\mathbf{q}_{4}}\left\langle \hat{c}_{\mathbf{q}_{3}}^{\dagger}\hat{c}_{\mathbf{q}_{1}}^{\dagger}\hat{c}_{\mathbf{q}_{4}}\hat{c}_{\mathbf{q}_{3}+\mathbf{q}_{2}-\mathbf{q}_{4}}\hat{c}_{\mathbf{q}+\mathbf{q}_{1}-\mathbf{q}_{2}}|t\right\rangle\times \nonumber \\
\fl & &\times\int_{-\infty}^{0}d\tau\:\mbox{e}^{\left[\varepsilon+i(\omega_{\mathbf{q}_{3}}+\omega_{\mathbf{q}_{2}}-\omega_{\mathbf{q}_{4}}-\omega_{\mathbf{q}_{3}+\mathbf{q}_{2}-\mathbf{q}_{4}})\right]\tau}+\nonumber \\
\fl & +4\hbar^{-2}&\sum_{\mathbf{q}_{1},\mathbf{q}_{2},\mathbf{q}_{3},\mathbf{q}_{4}}\mathcal{V}_{\mathbf{q},\mathbf{q}_{1},\mathbf{q}_{2}}\mathcal{V}_{\mathbf{q}_{3},\mathbf{q}_{4},\mathbf{q}_{1}}\left\langle \hat{c}_{\mathbf{q}_{3}}^{\dagger}\hat{c}_{\mathbf{q}_{4}}^{\dagger}\hat{c}_{\mathbf{q}_{2}}\hat{c}_{\mathbf{q}_{3}+\mathbf{q}_{4}-\mathbf{q}_{1}}\hat{c}_{\mathbf{q}+\mathbf{q}_{1}-\mathbf{q}_{2}}|t\right\rangle\times \nonumber \\
\fl & & \times\int_{-\infty}^{0}d\tau\:\mbox{e}^{\left[\varepsilon+i(\omega_{\mathbf{q}_{3}}+\omega_{\mathbf{q}_{4}}-\omega_{\mathbf{q}_{1}}-\omega_{\mathbf{q}_{3}+\mathbf{q}_{4}-\mathbf{q}_{1}})\right]\tau}-\nonumber \\
\fl & -4\hbar^{-2}&\sum_{\mathbf{q}_{1},\mathbf{q}_{2},\mathbf{q}_{3}}\mathcal{V}_{\mathbf{q},\mathbf{q}_{1},\mathbf{q}_{2}}\mathcal{V}_{\mathbf{q}_{2},\mathbf{q}+\mathbf{q}_{1}-\mathbf{q}_{2},\mathbf{q}_{3}}\left\langle \hat{c}_{\mathbf{q}_{1}}^{\dagger}\hat{c}_{\mathbf{q}_{3}}\hat{c}_{\mathbf{q}+\mathbf{q}_{1}-\mathbf{q}_{3}}|t\right\rangle\times \nonumber \\
\fl &  & \times\int_{-\infty}^{0}d\tau\:\mbox{e}^{\left[\varepsilon+i(\omega_{\mathbf{q}_{2}}+\omega_{\mathbf{q}+\mathbf{q}_{1}-\mathbf{q}_{2}}-\omega_{\mathbf{q}_{3}}-\omega_{\mathbf{q}+\mathbf{q}_{1}-\mathbf{q}_{3}})\right]\tau},\label{eq:J2_c_MM_explicit}\end{eqnarray}
where $\tau=t'-t$. Average values are calculated and then
expressed in terms of populations, amplitudes and pairs of magnons,
as discussed in the following Appendix D. The average value enclosed
in the last line, for example, gives \begin{eqnarray}
\fl\left\langle \hat{c}_{\mathbf{q}_{1}}^{\dagger}\hat{c}_{\mathbf{q}_{3}}\hat{c}_{\mathbf{q}+\mathbf{q}_{1}-\mathbf{q}_{3}}|t\right\rangle = & \left[\mathcal{N}_{\mathbf{q}_{1}}(t)-\left|\left\langle \hat{c}_{\mathbf{q}_{1}}|t\right\rangle \right|^{2}\right]\left(\delta_{\mathbf{q}_{3},\mathbf{q}_{1}}\left\langle \hat{c}_{\mathbf{q}}|t\right\rangle +\delta_{\mathbf{q}_{3},\mathbf{q}}\left\langle \hat{c}_{\mathbf{q}_{1}}|t\right\rangle \right)+\nonumber \\
\fl & +\left[\sigma_{\mathbf{q}_{3}}(t)-\left\langle \hat{c}_{\mathbf{q}_{3}}|t\right\rangle \left\langle \hat{c}_{-\mathbf{q}_{3}}|t\right\rangle \right]\delta_{\mathbf{q}_{1},\mathbf{q}}\left\langle \hat{c}_{\mathbf{q}}^{\dagger}|t\right\rangle +\nonumber \\
\fl & +\left\langle \hat{c}_{\mathbf{q}_{1}}^{\dagger}|t\right\rangle \left\langle \hat{c}_{\mathbf{q}_{3}}|t\right\rangle \left\langle \hat{c}_{\mathbf{q}+\mathbf{q}_{1}-\mathbf{q}_{3}}|t\right\rangle .\end{eqnarray}

The integration in time together with the limit of $\varepsilon\to+0$
produces the so-called retarded Heisenberg delta function, that is
\begin{equation}
\lim_{\varepsilon\to+0}\int_{-\infty}^{0}d\tau\:\mbox{e}^{\left[\varepsilon+i\omega\right]\tau}=\mbox{PV}\frac{1}{\omega}-i\pi\delta(\omega),\end{equation}
 where $\mbox{PV}$ stands for principal value. 

Applying this procedure to the all parts of $J_{c_{\mathbf{q}}}^{(2)}$
one obtains the evolution equation of the amplitudes, Eq. (\ref{eq:q_evolution_Js-1}).
Despite the extension of the obtained expressions, it is easy to note
that all terms possess linear dependence on the amplitudes, as shown
in the Eqs. (\ref{eq:J_0_c_q_apendice}) and (\ref{eq:J_0_c_q_apendice})
and because of the odd number of creation or annihilation operators
in the commutators of Eqs. (\ref{eq:J2_c_II})-(\ref{eq:J2_c_SR}).
Considering then a linear approximation and neglecting the self energy correction to the frequencies $\omega_\mathbf{q}$ we may write 

\begin{equation}
\frac{d}{dt}\left\langle \hat{c}_{\mathbf{q}}|t\right\rangle =-i\omega_{\mathbf{q}}\left\langle \hat{c}_{\mathbf{q}}|t\right\rangle -\Gamma_{\mathbf{q}}(t)\left\langle \hat{c}_{\mathbf{q}}|t\right\rangle ,\label{eq:amplitude_evolution_2-1}\end{equation}
 with frequency $\omega_\mathbf{q}$ and {\small \begin{eqnarray}
\vspace{-2cm}\Gamma_{\mathbf{q}}(t)& = 8\pi\hbar^{-2}\sum_{\mathbf{q}_{1},\mathbf{q}_{2},\mathbf{q}_{3}}\left|\mathcal{V}_{\mathbf{q},\mathbf{q}_{1},\mathbf{q}_{2}}\right|^{2}\left(\mathcal{N}_{\mathbf{q}_{2}}+\mathcal{N}_{\mathbf{q}+\mathbf{q}_{1}-\mathbf{q}_{2}}+1\right)\mathcal{N}_{\mathbf{q}_{1}}\times\nonumber \\
 & \qquad\qquad\times\delta(\omega_{\mathbf{q}}+\omega_{\mathbf{q}_{1}}-\omega_{\mathbf{q}_{2}}-\omega_{\mathbf{q}+\mathbf{q}_{1}-\mathbf{q}_{2}})-\nonumber \\
 & -8\pi\hbar^{-2}\sum_{\mathbf{q}_{1},\mathbf{q}_{2},\mathbf{q}_{3}}\left|\mathcal{V}_{\mathbf{q},\mathbf{q}_{1},\mathbf{q}_{2}}\right|^{2}\mathcal{N}_{\mathbf{q}_{2}}\mathcal{N}_{\mathbf{q}+\mathbf{q}_{1}-\mathbf{q}_{2}}\times\nonumber \\
 & \qquad\qquad\times\delta(\omega_{\mathbf{q}}+\omega_{\mathbf{q}_{1}}-\omega_{\mathbf{q}_{2}}-\omega_{\mathbf{q}+\mathbf{q}_{1}-\mathbf{q}_{2}})+\nonumber \\
 & +\pi\hbar^{-2}\sum_{\mathbf{k}\neq0}\left|\mathcal{F}_{\mathbf{q},\mathbf{k}}\right|^{2}\Big[\left(\mathcal{N}_{\mathbf{q}-\mathbf{k}}+\nu_{\mathbf{k}}+1\right)\delta(\Omega_{\mathbf{k}}+\omega_{\mathbf{q}-\mathbf{k}}-\omega_{\mathbf{q}})-\nonumber \\
 & \qquad\qquad-\left(\mathcal{N}_{\mathbf{q}-\mathbf{k}}-\nu_{-\mathbf{k}}\right)\delta(\Omega_{-\mathbf{k}}-\omega_{\mathbf{q}-\mathbf{k}}+\omega_{\mathbf{q}})\Big]+\nonumber \\
 & +4\pi\hbar^{-2}\sum_{\mathbf{k}\neq0}\left|\mathcal{L}_{\mathbf{q},\mathbf{k}}\right|^{2}\left(\mathcal{N}_{\mathbf{k}-\mathbf{q}}-\nu_{\mathbf{k}}\right)\delta(\Omega_{\mathbf{k}}-\omega_{\mathbf{q}}-\omega_{\mathbf{k}-\mathbf{q}})+\nonumber \\
 & +\hbar^{-2}\sum_{\mathbf{k}\neq0}\left|\mathcal{R}_{-\mathbf{q},\mathbf{k}}\right|^{2}\left(\nu_{\mathbf{k}}-\nu_{\mathbf{k}+\mathbf{q}}\right)\delta(\Omega_{\mathbf{k}+\mathbf{q}}-\Omega_{\mathbf{k}}-\omega_{\mathbf{q}})+\nonumber \\
 & +2\hbar^{-2}\sum_{\mathbf{k}\neq0}\left|\mathcal{R}_{\mathbf{q},\mathbf{k}}^{+}\right|^{2}\left(\nu_{-\mathbf{k}}+\nu_{\mathbf{q}+\mathbf{k}}+1\right)\delta(\Omega_{-\mathbf{k}}+\Omega_{\mathbf{q}+\mathbf{k}}-\omega_{\mathbf{q}})+\nonumber \\
 & +\pi\hbar^{-2}\sum_{\alpha,\mathbf{p}\neq0}\left|\mathcal{S}_{\alpha,\mathbf{q},\mathbf{p}}^{\parallel\mathrm{a}}\right|^{2}\left(\mathcal{N}_{\mathbf{q}-\mathbf{p}}+f_{\mathbf{p}}+1\right)\delta(\zeta_{\mathbf{p}}+\omega_{\mathbf{q}-\mathbf{p}}-\omega_{\mathbf{q}})-\nonumber \\
 & -\pi\hbar^{-2}\sum_{\alpha,\mathbf{p}\neq0}\left|\mathcal{S}_{\alpha,\mathbf{q},\mathbf{p}}^{\parallel\mathrm{a}}\right|^{2}\left(\mathcal{N}_{\mathbf{q}-\mathbf{p}}-f_{-\mathbf{p}}\right)\delta(\zeta_{-\mathbf{p}}-\omega_{\mathbf{q}-\mathbf{p}}+\omega_{\mathbf{q}})+\nonumber \\
 & +4\pi\hbar^{-2}\sum_{\alpha,\mathbf{p}\neq0}\left|\mathcal{S}_{\alpha,\mathbf{q},\mathbf{p}}^{\parallel\mathrm{b}}\right|^{2}\left(\mathcal{N}_{\mathbf{p}-\mathbf{q}}-f_{\mathbf{p}}\right)\delta(\zeta_{\mathbf{p}}-\omega_{\mathbf{q}}-\omega_{\mathbf{p}-\mathbf{q}}).\label{eq:gama}\end{eqnarray}
}{\small \par}

The pairs and populations of magnons equations of evolution are obtained
in an analogous proceeding. After calculating the collision integrals
the evolution equation of the pairs of magnons may be written as \begin{equation}
\frac{d}{dt}\sigma_{\mathbf{q}}(t)=-2i\omega_{\mathbf{q}}\,\sigma_{\mathbf{q}}-\Gamma_{\mathbf{q}}(t)\,\sigma_{\mathbf{q}}+\Lambda_{\mathbf{q}}(t).\label{eq:pares_evol}\end{equation}
 The first therm is associated with the frequency of precession of pairs,
$2\omega_\mathbf{q}$, and the second is a decay term ruled by the same
function $\Gamma_{\mathbf{q}}(t)$ (Eq. \ref{eq:gama}) related with
the amplitude decay. $\Lambda_{\mathbf{q}}(t)$ represent the non-linear
terms, \begin{equation}
\Lambda_{\mathbf{q}}(t)=\left.\frac{d\sigma_{\mathbf{q}}}{dt}\right|_{\mathrm{SL}}+\left.\frac{d\sigma_{\mathbf{q}}}{dt}\right|_{\mathrm{SR}}+A_{\mathbf{q}'}(t)+B_{\mathbf{q}'}(t) \end{equation}
The terms ${\displaystyle \left.\frac{d\sigma_{\mathbf{q}}}{dt}\right|_{\mathrm{SL}}}$
e ${\displaystyle \left.\frac{d\sigma_{\mathbf{q}}}{dt}\right|_{\mathrm{SR}}}$
are originated in spin-lattice and spin-radiation interactions and
depend exclusively on magnons populations; $A_{\mathbf{q}'}(t)$ are
non-linear combinations of pairs and populations; the last therm represent
the amplitude contributions to the pairs evolution.

Finally, the populations evolution equation is 
\small
\begin{eqnarray}
\fl & {\displaystyle \frac{d}{dt}}\mathcal{N}_{\mathbf{q}}(t)=\nonumber \\
\fl & = +\frac{8\pi}{\hbar^{2}}\sum_{\mathbf{q}'\neq-\mathbf{q}}\left|\mathcal{S}_{\mathbf{q},\mathbf{q}+\mathbf{q}'}^{\parallel\mathrm{b}}\right|^{2}\left\{ (1+\mathcal{N}_{\mathbf{q}}+\mathcal{N}_{\mathbf{q}'})f_{\mathbf{q}'+\mathbf{q}}^{\mathrm{S}}\right\} \delta(\omega_{\mathbf{q}}+\omega_{\mathbf{q}'}-\zeta_{\mathbf{q}+\mathbf{q}'})+ & [\mathfrak{S}_{\mathbf{q}}(t)]\nonumber \\
\fl & +{\displaystyle \frac{8\pi}{\hbar^{2}}\sum_{\mathbf{q}'\neq-\mathbf{q}}}\left|\mathcal{S}_{\mathbf{q},\mathbf{q}+\mathbf{q}'}^{\parallel\mathrm{b}}\right|^{2}\left\{ (\mathcal{N}_{\mathbf{q}'}+1)(\mathcal{N}_{\mathbf{q}}+1)f_{\mathbf{q}'+\mathbf{q}}^{\mathrm{T}}-\mathcal{N}_{\mathbf{q}'}\mathcal{N}_{\mathbf{q}}(f_{\mathbf{q}'+\mathbf{q}}^{\mathrm{T}}+1)\right\} \delta(\omega_{\mathbf{q}}+\omega_{\mathbf{q}'}-\zeta_{\mathbf{q}+\mathbf{q}'})- & [\mathfrak{R}_{\mathbf{q}}(t)]\nonumber \\
\fl & -{\displaystyle \frac{1}{\tau_{\mathbf{q}}}}\left[\mathcal{N}_{\mathbf{q}}-\mathcal{N}_{\mathbf{q}}^{(0)}\right]+ & [L_{\mathbf{q}}(t)]\nonumber \\
\fl & +{\displaystyle \frac{8\pi}{\hbar^{2}}\sum_{\mathbf{q}'\neq-\mathbf{q}}}\left|\mathrm{\mathcal{L}}_{\mathbf{q},\mathbf{q}+\mathbf{q}'}\right|^{2}\left\{ (\mathcal{N}_{\mathbf{q}'}+1)(\mathcal{N}_{\mathbf{q}}+1)\nu_{\mathbf{q}'+\mathbf{q}}-\mathcal{N}_{\mathbf{q}'}\mathcal{N}_{\mathbf{q}}(\nu_{\mathbf{q}'+\mathbf{q}}+1)\right\} \delta(\omega_{\mathbf{q}}+\omega_{\mathbf{q}'}-\Omega_{\mathbf{q}+\mathbf{q}'})+ & [\mathfrak{L}_{\mathbf{q}}(t)]\nonumber \\
\fl & \vspace{-1cm}\begin{array}{l}
+{\displaystyle \frac{2\pi}{\hbar^{2}}\sum_{\mathbf{q}'\neq\mathbf{q}}}\left|\mathcal{F}_{\mathbf{q},\mathbf{q}-\mathbf{q}'}\right|^{2}\left\{ \mathcal{N}_{\mathbf{q}'}(\mathcal{N}_{\mathbf{q}}+1)(\nu_{\mathbf{q}'-\mathbf{q}}+1)-(\mathcal{N}_{\mathbf{q}'}+1)\mathcal{N}_{\mathbf{q}}\nu_{\mathbf{q}'-\mathbf{q}}\right\} \delta(\omega_{\mathbf{q}'}-\omega_{\mathbf{q}}-\Omega_{\mathbf{q}'-\mathbf{q}})+\\
+{\displaystyle \frac{2\pi}{\hbar^{2}}\sum_{\mathbf{q}'\neq\mathbf{q}}}\left|\mathcal{F}_{\mathbf{q},\mathbf{q}-\mathbf{q}'}\right|^{2}\left\{ (\mathcal{N}_{\mathbf{q}}+1)\mathcal{N}_{\mathbf{q}'}\nu_{\mathbf{q}-\mathbf{q}'}-\mathcal{N}_{\mathbf{q}}(\mathcal{N}_{\mathbf{q}'}+1)(\nu_{\mathbf{q}-\mathbf{q}'}+1)\right\} \delta(\omega_{\mathbf{q}'}-\omega_{\mathbf{q}}+\Omega_{\mathbf{q}-\mathbf{q}'})+\end{array} & [\mathfrak{F}_{\mathbf{q}}(t)]\nonumber \\
\fl & +{\displaystyle \frac{16\pi}{\hbar^{2}}\sum_{\mathbf{q}_{1},\mathbf{q}_{2},\mathbf{q}_{3}}}\begin{array}{c}
\left|\mathcal{V}_{\mathbf{q},\mathbf{q}_{1},\mathbf{q}_{2}}\right|^{2}\left\{ (\mathcal{N}_{\mathbf{q}}+1)(\mathcal{N}_{\mathbf{q}_{1}}+1)\mathcal{N}_{\mathbf{q}_{2}}\mathcal{N}_{\mathbf{q}_{3}}-\mathcal{N}_{\mathbf{q}}\mathcal{N}_{\mathbf{q}_{1}}(\mathcal{N}_{\mathbf{q}_{2}}+1)(\mathcal{N}_{\mathbf{q}_{3}}+1)\right\} \times\\
\times\delta(\omega_{\mathbf{q}}+\omega_{\mathbf{q}_{1}}-\omega_{\mathbf{q}_{2}}-\omega_{\mathbf{q}_{3}})\delta_{\mathbf{q}_{3},\mathbf{q}+\mathbf{q}_{1}-\mathbf{q}_{2}}\end{array}+ & [\mathfrak{M}_{\mathbf{q}}(t)]\nonumber \\
\fl & +\sum_{\mathbf{q}'\neq\mathbf{q}}C_{\mathbf{q}'}\left(\left\langle \hat{c}_{\mathbf{q}'}|t\right\rangle ,\left\langle \hat{c}_{\mathbf{q}'}^{\dagger}|t\right\rangle ,\sigma_{\mathbf{q}'},\sigma_{\mathbf{q}'}^{*},t\right), & [\mathfrak{A}_{\mathbf{q}}(t)]\nonumber \\
\fl & &\label{eq:N_q_evol}\end{eqnarray}
\normalsize
where $f_{\mathbf{p}^{\mathrm{S}}}$ and $f_{\mathbf{p}^{\mathrm{T}}}$ are the average photon populations due to the source and black-body radiation respectively, and $C_{\mathbf{q}'}\left(\left\langle \hat{c}_{\mathbf{q}'}|t\right\rangle ,\left\langle \hat{c}_{\mathbf{q}'}^{\dagger}|t\right\rangle ,\sigma_{\mathbf{q}'},\sigma_{\mathbf{q}'}^{*},t\right)$ are the contributions from amplitudes and pairs to the evolution. In compact form, as in Eq. (\ref{eq:population_evolution_2}),
\begin{equation}
\frac{d}{dt}\mathcal{N}_{\mathbf{q}}(t)=\mathfrak{S}_{\mathbf{q}}(t)+\mathfrak{R}_{\mathbf{q}}(t)+L_{\mathbf{q}}(t)+\mathfrak{L}_{\mathbf{q}}(t)+\mathfrak{F}_{\mathbf{q}}(t)+\mathfrak{M}_{\mathbf{q}}(t)+\mathfrak{A}_{\mathbf{q}}(t)\,.\label{eq:population_evolution_2-1}\end{equation}

\section{Average Values}

The macrovariables of set in Eq. (\ref{eq:macrovariables}) are the
average of the set of microvariables in Eq. (\ref{eq:microdynamical_variables_3})
weighted with the auxiliary statistical operator \small\begin{eqnarray}
\fl\hat{\bar{\varrho}}(t,0)=\frac{\exp\left\{ -{\displaystyle \sum_{\mathbf{q}}}\left[F_{\mathbf{q}}(t)\hat{c}_{\mathbf{q}}^{\dagger}\hat{c}_{\mathbf{q}}+\phi_{\mathbf{q}}(t)\hat{c}_{\mathbf{q}}+\phi_{\mathbf{q}}^{*}(t)\hat{c}_{\mathbf{q}}^{\dagger}+\varphi_{\mathbf{q}}(t)\hat{c}_{\mathbf{q}}\hat{c}_{-\mathbf{q}}+\varphi_{\mathbf{q}}^{*}(t)\hat{c}_{\mathbf{q}}^{\dagger}\hat{c}_{-\mathbf{q}}^{\dagger}\right]\right\} }{\mbox{Tr}\exp\left\{ -{\displaystyle \sum_{\mathbf{q}}}\left[F_{\mathbf{q}}(t)\hat{c}_{\mathbf{q}}^{\dagger}\hat{c}_{\mathbf{q}}+\phi_{\mathbf{q}}(t)\hat{c}_{\mathbf{q}}+\phi_{\mathbf{q}}^{*}(t)\hat{c}_{\mathbf{q}}^{\dagger}+\varphi_{\mathbf{q}}(t)\hat{c}_{\mathbf{q}}\hat{c}_{-\mathbf{q}}+\varphi_{\mathbf{q}}^{*}(t)\hat{c}_{\mathbf{q}}^{\dagger}\hat{c}_{-\mathbf{q}}^{\dagger}\right]\right\} }\:.\nonumber \\
\end{eqnarray} 
\normalsize
This averages can be done through diagonalization of $\hat{\bar{\rho}}(t,0)$.
Using the following transformation \begin{eqnarray}
\hat{c}_{\mathbf{q}}=\iota_{\mathbf{q}}\cm_{\mathbf{q}}+\kappa_{\mathbf{q}}\cm_{-\mathbf{q}}^{\dagger}+\lambda_{\mathbf{q}},\nonumber \\
\hat{c}_{-\mathbf{q}}^{\dagger}=\iota_{\mathbf{q}}\cm_{-\mathbf{q}}^{\dagger}+\kappa_{\mathbf{q}}\cm_{\mathbf{q}}+\lambda_{\mathbf{-q}}^{*},\end{eqnarray}
 where $\cm_{\mathbf{q}}^{\dagger}$ and $\cm_{\mathbf{q}}$ are bosonic
operators and $\lambda_{\mathbf{q}}$, $\iota_{\mathbf{q}}$ and $\kappa_{\mathbf{q}}$
functions of $\mathbf{q}$ such that \begin{eqnarray}
1\: & =\left[\hat{c}_{\mathbf{q}},\hat{c}_{\mathbf{q}}^{\dagger}\right]=\nonumber \\
 & =\left[\iota_{\mathbf{q}}\cm_{\mathbf{q}}+\kappa_{\mathbf{q}}\cm_{-\mathbf{q}}^{\dagger},\iota_{\mathbf{q}}^{*}\cm_{\mathbf{q}}^{\dagger}+\kappa_{\mathbf{q}}^{*}\cm_{-\mathbf{q}}\right]=\nonumber \\
 & =\left|\iota_{\mathbf{q}}\right|^{2}\left[\cm_{\mathbf{q}},\cm_{\mathbf{q}}^{\dagger}\right]+\left|\kappa_{\mathbf{q}}\right|^{2}\left[\cm_{-\mathbf{q}}^{\dagger},\cm_{-\mathbf{q}}\right]=\left|\iota_{\mathbf{q}}\right|^{2}-\left|\kappa_{\mathbf{q}}\right|^{2},\end{eqnarray}
 it can be shown that

\begin{eqnarray}
\bar{\varrho}(t,0)=\frac{\exp\left\{ -{\displaystyle \sum_{\mathbf{q}}}F_{\mathbf{q}}'(t)\cm_{\mathbf{q}}^{\dagger}\cm_{\mathbf{q}}\right\} }{\mbox{Tr}\exp\left\{ -{\displaystyle \sum_{\mathbf{q}}}F_{\mathbf{q}}'(t)\cm_{\mathbf{q}}^{\dagger}\cm_{\mathbf{q}}\right\} }\:,\end{eqnarray}
 if \begin{equation}
\lambda_{\mathbf{q}}=-\frac{\phi_{\mathbf{q}}^{*}(t)}{F_{\mathbf{q}}(t)+\left[\varphi_{\mathbf{q}}^{*}(t)+\varphi_{-\mathbf{q}}^{*}(t)\right]}\:,\label{eq:diago_lambda}\end{equation}

\begin{eqnarray}
\iota_{\mathbf{q}}=\sqrt{\frac{F_{\mathbf{q}}(t)+F_{\mathbf{q}}'(t)}{2F_{\mathbf{q}}'(t)}}\:, \qquad \kappa_{\mathbf{q}}=\sqrt{\frac{F_{\mathbf{q}}(t)-F_{\mathbf{q}}'(t)}{2F_{\mathbf{q}}'(t)}}\:,\label{eq:diago_iota_kappa}\end{eqnarray}
 and \begin{equation}
F_{\mathbf{q}}'(t)=\sqrt{F_{\mathbf{q}}^{2}(t)-\left|\varphi_{\mathbf{q}}(t)+\varphi_{-\mathbf{q}}(t)\right|^{2}}.\label{eq:diago_F}\end{equation}

Using this transformation we show that the amplitudes are   \begin{eqnarray}
\fl\left\langle \hat{c}_{\mathbf{q}}|t\right\rangle & = \mbox{Tr}\left\{ \hat{c}_{\mathbf{q}}\,\hat{\bar{\varrho}}(t,0)\right\} =\mbox{Tr}\left\{ \left(\iota_{\mathbf{q}}\cm_{\mathbf{q}}+\kappa_{\mathbf{q}}\cm_{-\mathbf{q}}^{\dagger}+\lambda_{\mathbf{q}}\right)\,\hat{\bar{\varrho}}(t,0)\right\} =\nonumber \\
\fl& = \iota_{\mathbf{q}}\mbox{Tr}\left\{ \cm_{\mathbf{q}}\,\hat{\bar{\varrho}}(t,0)\right\} +\kappa_{\mathbf{q}}\mbox{Tr}\left\{ \cm_{-\mathbf{q}}^{\dagger}\,\hat{\bar{\varrho}}(t,0)\right\} +\lambda_{\mathbf{q}}\mbox{Tr}\left\{ \,\hat{\bar{\varrho}}(t,0)\right\} =\lambda_{\mathbf{q}},\label{eq:aniquila_medio}\end{eqnarray}

\begin{eqnarray}
\fl\left\langle \hat{c}_{\mathbf{q}}^{\dagger}|t\right\rangle & = \mbox{Tr}\left\{ \hat{c}_{\mathbf{q}}^{\dagger}\,\hat{\bar{\varrho}}(t,0)\right\} =\mbox{Tr}\left\{ \left(\iota_{\mathbf{q}}^{*}\cm_{\mathbf{q}}^{\dagger}+\kappa_{\mathbf{q}}^{*}\cm_{-\mathbf{q}}+\lambda_{\mathbf{q}}^{*}\right)\,\hat{\bar{\varrho}}(t,0)\right\} =\nonumber \\
\fl& = \iota_{\mathbf{q}}^{*}\mbox{Tr}\left\{ \cm_{\mathbf{q}}^{\dagger}\,\hat{\bar{\varrho}}(t,0)\right\} +\kappa_{\mathbf{q}}^{*}\mbox{Tr}\left\{ \cm_{-\mathbf{q}}\,\hat{\bar{\varrho}}(t,0)\right\} +\lambda_{\mathbf{q}}^{*}\mbox{Tr}\left\{ \hat{\bar{\varrho}}(t,0)\right\} =\lambda_{\mathbf{q}}^{*},\label{eq:cria_medio}\end{eqnarray}
 where we used the normalization condition $\mbox{Tr}\left\{ \hat{\bar{\varrho}}(t,0)\right\} =1$
and that the first two contributions on the right are null. Average
values for two magnon operators are \begin{eqnarray}
\fl\left\langle \hat{c}_{\mathbf{q}_{a}}\hat{c}_{\mathbf{q}_{b}}|t\right\rangle = \mbox{Tr}\left\{ \hat{c}_{\mathbf{q}_{a}}\hat{c}_{\mathbf{q}_{b}}\,\hat{\bar{\varrho}}(t,0)\right\} \nonumber \\
\fl=\mbox{Tr}\left\{ \left(\iota_{\mathbf{q}_{a}}\cm_{\mathbf{q}_{a}}+\kappa_{\mathbf{q}_{a}}\cm_{-\mathbf{q}_{a}}^{\dagger}+\lambda_{\mathbf{q}_{a}}\right)\left(\iota_{\mathbf{q}_{b}}\cm_{\mathbf{q}_{b}}+\kappa_{\mathbf{q}_{b}}\cm_{-\mathbf{q}_{b}}^{\dagger}+\lambda_{\mathbf{q}_{b}}\right)\,\hat{\bar{\varrho}}(t,0)\right\} =\nonumber \\
\fl= \iota_{\mathbf{q}_{a}}\iota_{\mathbf{q}_{b}}\mbox{Tr}\left\{ \cm_{\mathbf{q}_{a}}\cm_{\mathbf{q}_{b}}\,\hat{\bar{\varrho}}(t,0)\right\} +\iota_{\mathbf{q}_{a}}\kappa_{\mathbf{q}_{b}}\mbox{Tr}\left\{ \cm_{\mathbf{q}_{a}}\cm_{-\mathbf{q}_{b}}^{\dagger}\,\hat{\bar{\varrho}}(t,0)\right\} +\iota_{\mathbf{q}_{a}}\lambda_{\mathbf{q}_{b}}\mbox{Tr}\left\{ \cm_{\mathbf{q}_{a}}\,\hat{\bar{\varrho}}(t,0)\right\} +\nonumber \\
\fl+\kappa_{\mathbf{q}_{a}}\iota_{\mathbf{q}_{b}}\mbox{Tr}\left\{ \cm_{-\mathbf{q}_{a}}^{\dagger}\cm_{\mathbf{q}_{b}}\,\hat{\bar{\varrho}}(t,0)\right\} +\kappa_{\mathbf{q}_{a}}\kappa_{\mathbf{q}_{b}}\mbox{Tr}\left\{ \cm_{-\mathbf{q}_{a}}^{\dagger}\cm_{-\mathbf{q}_{b}}^{\dagger}\,\hat{\bar{\varrho}}(t,0)\right\} +\kappa_{\mathbf{q}_{a}}\lambda_{\mathbf{q}_{b}}\mbox{Tr}\left\{ \cm_{-\mathbf{q}_{a}}^{\dagger}\,\hat{\bar{\varrho}}(t,0)\right\} +\nonumber \\
\fl+\lambda_{\mathbf{q}_{a}}\iota_{\mathbf{q}_{b}}\mbox{Tr}\left\{ \cm_{\mathbf{q}_{b}}\,\hat{\bar{\varrho}}(t,0)\right\} +\lambda_{\mathbf{q}_{a}}\kappa_{\mathbf{q}_{b}}\mbox{Tr}\left\{ \cm_{-\mathbf{q}_{b}}^{\dagger}\,\hat{\bar{\varrho}}(t,0)\right\} +\lambda_{\mathbf{q}_{a}}\lambda_{\mathbf{q}_{b}}\mbox{Tr}\left\{ \hat{\bar{\varrho}}(t,0)\right\} =\nonumber \\
\fl= \iota_{\mathbf{q}_{a}}\kappa_{\mathbf{q}_{b}}\mbox{Tr}\left\{ \cm_{\mathbf{q}_{a}}\cm_{-\mathbf{q}_{b}}^{\dagger}\,\hat{\bar{\varrho}}(t,0)\right\} +\kappa_{\mathbf{q}_{a}}\iota_{\mathbf{q}_{b}}\mbox{Tr}\left\{ \cm_{-\mathbf{q}_{a}}^{\dagger}\cm_{\mathbf{q}_{b}}\,\hat{\bar{\varrho}}(t,0)\right\} +\lambda_{\mathbf{q}_{a}}\lambda_{\mathbf{q}_{b}}\mbox{Tr}\left\{ \hat{\bar{\varrho}}(t,0)\right\} =\nonumber \\
\fl= \left\{ \frac{\iota_{\mathbf{q}_{a}}\kappa_{\mathbf{q}_{b}}}{1-\mbox{e}^{-F'_{\mathbf{q}_{a}}}}+\frac{\kappa_{\mathbf{q}_{a}}\iota_{\mathbf{q}_{b}}}{\mbox{e}^{F'_{\mathbf{q}_{b}}}-1}\right\} \delta_{\mathbf{q}_{a},-\mathbf{q}_{b}}+\left\langle \hat{c}_{\mathbf{q}_{a}}|t\right\rangle \left\langle \hat{c}_{\mathbf{q}_{b}}|t\right\rangle ,\end{eqnarray}
 after using Eqs. (\ref{eq:aniquila_medio}) and (\ref{eq:cria_medio}).
For $\mathbf{q}_{a}=\mathbf{q}_{b}\equiv\mathbf{q}$ we have that
\begin{equation}
\sigma_{\mathbf{q}}(t)=\left\langle \hat{c}_{\mathbf{q}}\hat{c}_{-\mathbf{q}}|t\right\rangle =\frac{\iota_{\mathbf{q}}\kappa_{-\mathbf{q}}}{1-\mbox{e}^{-F'_{\mathbf{q}}}}+\frac{\kappa_{\mathbf{q}}\iota_{-\mathbf{q}}}{\mbox{e}^{F'_{-\mathbf{q}}}-1}+\left\langle \hat{c}_{\mathbf{q}}|t\right\rangle \left\langle \hat{c}_{\mathbf{q}}|t\right\rangle .\end{equation}

\begin{eqnarray}
\fl\left\langle \hat{c}_{\mathbf{q}_{a}}^{\dagger}\hat{c}_{\mathbf{q}_{b}}|t\right\rangle = \mbox{Tr}\left\{ \hat{c}_{\mathbf{q}_{a}}^{\dagger}\hat{c}_{\mathbf{q}_{b}}\,\hat{\bar{\varrho}}(t,0)\right\} = \nonumber \\
\fl=\mbox{Tr}\left\{ \left(\iota_{\mathbf{q}_{a}}^{*}\cm_{\mathbf{q}_{a}}^{\dagger}+\kappa_{\mathbf{q}_{a}}^{*}\cm_{-\mathbf{q}_{a}}+\lambda_{\mathbf{q}_{a}}^{*}\right)\left(\iota_{\mathbf{q}_{b}}\cm_{\mathbf{q}_{b}}+\kappa_{\mathbf{q}_{b}}\cm_{-\mathbf{q}_{b}}^{\dagger}+\lambda_{\mathbf{q}_{b}}\right)\,\hat{\bar{\varrho}}(t,0)\right\} =\nonumber \\
\fl= \iota_{\mathbf{q}_{a}}^{*}\iota_{\mathbf{q}_{b}}\mbox{Tr}\left\{ \cm_{\mathbf{q}_{a}}^{\dagger}\cm_{\mathbf{q}_{b}}\,\hat{\bar{\varrho}}(t,0)\right\} +\iota_{\mathbf{q}_{a}}^{*}\kappa_{\mathbf{q}_{b}}\mbox{Tr}\left\{ \cm_{\mathbf{q}_{a}}^{\dagger}\cm_{-\mathbf{q}_{b}}^{\dagger}\,\hat{\bar{\varrho}}(t,0)\right\} +\iota_{\mathbf{q}_{a}}^{*}\lambda_{\mathbf{q}_{b}}\mbox{Tr}\left\{ \cm_{\mathbf{q}_{a}}^{\dagger}\,\hat{\bar{\varrho}}(t,0)\right\} +\nonumber \\
\fl+\kappa_{\mathbf{q}_{a}}^{*}\iota_{\mathbf{q}_{b}}\mbox{Tr}\left\{ \cm_{-\mathbf{q}_{a}}\cm_{\mathbf{q}_{b}}\,\hat{\bar{\varrho}}(t,0)\right\} +\kappa_{\mathbf{q}_{a}}^{*}\kappa_{\mathbf{q}_{b}}\mbox{Tr}\left\{ \cm_{-\mathbf{q}_{a}}\cm_{-\mathbf{q}_{b}}^{\dagger}\,\hat{\bar{\varrho}}(t,0)\right\} +\kappa_{\mathbf{q}_{a}}^{*}\lambda_{\mathbf{q}_{b}}\mbox{Tr}\left\{ \cm_{-\mathbf{q}_{a}}\,\hat{\bar{\varrho}}(t,0)\right\} +\nonumber \\
\fl+\lambda_{\mathbf{q}_{a}}^{*}\iota_{\mathbf{q}_{b}}\mbox{Tr}\left\{ \cm_{\mathbf{q}_{b}}\,\hat{\bar{\varrho}}(t,0)\right\} +\lambda_{\mathbf{q}_{a}}^{*}\kappa_{\mathbf{q}_{b}}\mbox{Tr}\left\{ \cm_{-\mathbf{q}_{b}}^{\dagger}\,\hat{\bar{\varrho}}(t,0)\right\} +\lambda_{\mathbf{q}_{a}}^{*}\lambda_{\mathbf{q}_{b}}\mbox{Tr}\left\{ \hat{\bar{\varrho}}(t,0)\right\} =\nonumber \\
\fl= \iota_{\mathbf{q}_{a}}^{*}\iota_{\mathbf{q}_{b}}\mbox{Tr}\left\{ \cm_{\mathbf{q}_{a}}^{\dagger}\cm_{\mathbf{q}_{b}}\,\hat{\bar{\varrho}}(t,0)\right\} +\kappa_{\mathbf{q}_{a}}^{*}\kappa_{\mathbf{q}_{b}}\mbox{Tr}\left\{ \cm_{-\mathbf{q}_{a}}\cm_{-\mathbf{q}_{b}}^{\dagger}\,\hat{\bar{\varrho}}(t,0)\right\} +\lambda_{\mathbf{q}_{a}}^{*}\lambda_{\mathbf{q}_{b}}\mbox{Tr}\left\{ \hat{\bar{\varrho}}(t,0)\right\} =\nonumber \\
\fl= \left\{ \frac{\left|\iota_{\mathbf{q}_{a}}\right|^{2}}{\mbox{e}^{F'_{\mathbf{q}_{a}}}-1}+\frac{\left|\kappa_{-\mathbf{q}_{a}}\right|^{2}}{1-\mbox{e}^{-F'_{-\mathbf{q}_{a}}}}\right\} \delta_{\mathbf{q}_{a},\mathbf{q}_{b}}+\left\langle \hat{c}_{\mathbf{q}_{a}}^{\dagger}|t\right\rangle \left\langle \hat{c}_{\mathbf{q}_{b}}|t\right\rangle ,\end{eqnarray}
 and, if $\mathbf{q}_{a}=\mathbf{q}_{b}\equiv\mathbf{q}$,

\begin{equation}
\left\langle \hat{c}_{\mathbf{q}}^{\dagger}\hat{c}_{\mathbf{q}}|t\right\rangle =\mathcal{N}_{\mathbf{q}}(t)=\frac{\left|\iota_{\mathbf{q}}\right|^{2}}{\mbox{e}^{F'_{\mathbf{q}}}-1}+\frac{\left|\kappa_{-\mathbf{q}}\right|^{2}}{1-\mbox{e}^{-F'_{-\mathbf{q}}}}+\left|\left\langle \hat{c}_{\mathbf{q}}|t\right\rangle \right|^{2}.\label{eq:populacao(intensivas)_apendice}\end{equation}

Average values of more than two magnons operators are calculated in
analogous form.


\begin{thebibliography}{68}

\bibitem{demokritov2006}S.O. Demokritov et al., \emph{Bose-Einstein
condensation of quasi-equilibrium magnons at room temperature under
pumping}, Nature \textbf{443}, 430-433 (2006).

\bibitem{demidov2008}V.E. Demidov et al., \emph{Thermalization of a 
Parametrically Driven Magnon Gas Leading to Bose-Einstein Condensation},
Phys. Rev. Lett. \textbf{99}, 037205 (2007).

\bibitem{tupitsyn2008}I.S. Tupitsyn, P.C.E. Stamp and A.L. Burin,
Phys. Rev. Lett. \textbf{100}, 257202 (2008).

\bibitem{rezende2009}S. M. Rezende, Phys. Rev. B \textbf{79}, 174411,
(2009).

\bibitem{malomed2010}B.A. Malomed, O. Dzyapko, V.E. Demidov and S.
O. Demokritov, Phys. Rev. B \textbf{81}, 024418, (2010).

\bibitem{leggett2001}A.J. Leggett, \emph{Bose-Einstein condensation
in the alkali gases: Some fundamental concepts}, Rev. Mod. Phys. \textbf{73},
307 (2001).

\bibitem{pitaevskii2003} L.P. Pitaevskii and S. Stringari, \emph{Bose-Einstein
Condensation} (Clarendon, Oxford, UK, 2003).

\bibitem{snoke2010}D. Snoke and P. Littlewood, \emph{Polariton condensates},
Phys. Today \textbf{63}(8), 42 (2010).

\bibitem{snoke2006}D. Snoke, \emph{Coherent Questions}, Nature \textbf{443},
403 (2006).

\bibitem{phonon}J.C. Vaissiere et al., \emph{Numerical solution
of coupled steady-state hot-phonons-hot-electron Boltzmann equation
in InP}, Phys. Rev. B \textbf{46}, 13082 (1992).

\bibitem{froehlich1970}H. Fr\"{o}hlich, \emph{Long Range Coherence and
the Action of Enzymes}, Nature \textbf{228}, 1093-1093 (1970).

\bibitem{froehlich1980}H. Fr\"{o}hlich, \emph{The Biological Effects
of Microwaves and Related Questions}, in Adv. Electronics Electron
Phys. Vol. \textbf{53}, 88-192 (Academic, New York, USA, 1980).

\bibitem{mesquita1993}M.V. Mesquita, A.R. Vasconcellos and R. Luzzi,
\emph{Selective amplification of coherent polar vibrations in biopolymers},
Phys Rev. E \textbf{48}, 4049-4059 (1993).

\bibitem{fonseca2000}A.F. Fonseca, M.V. Mesquita, A.R. Vasconcellos
and R. Luzzi, \emph{Informational-statistical thermodynamics of a
complex system}, J. Chem. Phys. \textbf{112}(9) 3967-3979 (2000).

\bibitem{hyland1998}G.J. Hyland, \emph{Coherent GHz and THz Excitations
in Active Biosystems, and their Implications, in The Future of Medical
Diagnostics?} Matra Marconi UK, Directorate of Science, Internal Report
25.06.98, Portsmouth, UK (1998).

\bibitem{penrose1994}D. Penrose, \emph{Shadows of the Mind} (Oxford
Univ. Press, Oxford, UK, 1994).

\bibitem{lu1994} J. Lu, Z. Hehong and J.F. Greenleaf, \emph{Biomedical
ultrasound beam forming}, Ultrasound Med. Biol. \textbf{20}, 403-428
(1994).

\bibitem{mesquita1998}M.V. Mesquita, A.R. Vasconcellos and R. Luzzi,
\emph{Solitons in Highly Excited Matter: Dissipative Thermodynamics
and Supersonic Effects}, Phys. Rev. E \textbf{58}, 7913 (1998).

\bibitem{mysyrowics1996}A. Mysyrowicz, E. Benson and E. Fortin, \emph{Directed
Beams of Excitons Produced by Stimulated Scattering}, Phys. Rev. Lett.
\textbf{77}, 896 (1996).

\bibitem{mesquita2000}M.V. Mesquita, A.R. Vasconcellos and R. Luzzi,
\emph{{}``Excitoner\textquotedblright{}: Stimulated amplification
and propagation of excitons beams}, Europhys. Lett. \textbf{49}, 637-643
(2000).

\bibitem{vannucchi2010}F.S. Vannucchi, A.R. Vasconcellos and R.
Luzzi, \emph{Nonequilibrium Bose-Einstein condensation of hot magnons},
Phys. Rev. B \textbf{82}(14), 140404 (2010).

\bibitem{demodov2011}V.E. Demidov et al., \emph{Monochromatic microwave
radiation from the system of strongly excited magnon}, Appl. Phys
Lett. \textbf{92}, 162510 (2008).

\bibitem{ma2011}F.S. Ma et al., \emph{Micromagnetic study of spin
wave propagation in bicomponent magnonic component crystal waveguides},
Appl. Phys. Lett. \textbf{98}, 153107 (2011).

\bibitem{kent2006}A.J. Kent et al., \emph{Acoustic Phonon Emission
from a Weakly Coupled Superlattice}, Phys. Rev. Lett. \textbf{96},
215504 (2006).

\bibitem{rodrigues2011}C.G. Rodrigues, A.R. Vasconcellos and R.
Luzzi, \emph{Drifting Electron Excitation of Acoustic Phonons via
Piezoelectric Interaction}, as yet unpublished.

\bibitem{rodrigues2010}C.G. Rodrigues, A.R. Vasconcellos and R.
Luzzi, \emph{Evolution kinetics of nonequilibrium longitudinal-optical
phonons generated by drifting electrons in III-nitrides: longitudinal-optical-phonon
resonance}, J. Appl. Phys. \textbf{108}, 033716 (2010).

\bibitem{komirenko2000}S.M. Komirenko et al., \emph{Coherent optical
phonon generation by the electric current in quantum wells}, Appl.
Phys. Lett. \textbf{77}(25), 4178-4180 (2000).

\bibitem{keffer1966}F. Keffer, \emph{Spin} \emph{Waves in Handbuch
der Physik} XVIII/2 S. Fl\"{u}gge, Ed. (Springer, Berlin, Germany, 1966).

\bibitem{akhiezer1968}A.I. Akhiezer, V.G. Bar'yakhtar and S.V.
Peletminskii, \emph{Spin Waves} (North Holland, Amsterdam, The Netherlands,
1968).

\bibitem{white1983}R. M. White, \emph{Quantum Theory of Magnetism},
volume 32 of Springer Series in Solid-State Sciences (Springer-Verlag,
Berlin, Germany, 1983). 

\bibitem{luzzilivro2002}R. Luzzi, A.R. Vasconcellos and J.G. Ramos,
\emph{Predictive Statistical Mechanics: A Nonequilibrium Statistical
Ensemble Formalism} (Kluwer Academic, Dordrecht, The Netherlands,
2002): description in terms of the variational approach.

\bibitem{luzzi2006}R. Luzzi, A.R. Vasconcellos and J.G. Ramos,
\emph{The Theory of Irreversible processes: A Nonequilibrium Statistical
Ensemble Formalism}, Rivista del Nuovo Cimento \textbf{29}(2), 1-85,
(2006): alternative description in terms of a heuristic approach.

\bibitem{zubarev1996}D. N. Zubarev, V. Morosov and G. R\"{o}pke, \emph{Statistical
Mechanics of Nonequilibrium Processes}, Vols. 1 and 2 (Akademie Verlag,
Berlin, Germany, 1996 and 1997).

\bibitem{kuzensky2009}A.L. Kuzemsky, \emph{Statistical Mechanics
and the Physics of Many-Particle Model Systems}, Phys. Part. Nuclei
\textbf{40}, 949 (2009).

\bibitem{akhiezer1981}A.I. Akhiezer and S. V. Peletminskii, \emph{Methods
of Statistical Physics} (Pergamon, Oxford, UK, 1981).

\bibitem{mclennan1963}J.A. McLennan, \emph{Statistical Theory of
Transport Processes}, in Advances in Chemical Physics, Vol. \textbf{5},
261-317 (Academic, New York, USA, 1963).

\bibitem{kalos2007}M.H. Kalos, P.A. Whitlock, \emph{Monte Carlo
Methods} (Wiley Interscience, New York, USA, 2007).

\bibitem{frenkel2002}D. Frenkel, B. Smit, \emph{Understanding Molecular
Simulation} (Academic, New York, USA, 2002).

\bibitem{alder1987}B.J. Alder and D.J. Tildesley, \emph{Computer
Simulation of Liquids} (Oxford Univ. Press, New York, USA, 1987).

\bibitem{lauck1990}L. Lauck, A.R. Vasconcellos and R. Luzzi, \emph{A
Nonlinear Quantum Transport Theory}, Physica A \textbf{168}, 789-819
(1990).

\bibitem{kuzensky2007}A.L. Kuzemsky, \emph{Theory of Transport Processes
and the Method of the Nonequilibrium Statistical Operator}, Int. J.
Mod. Phys. B \textbf{21}(17), 2821 (2007).

\bibitem{madureira1998}A.J. Madureira, L. Lauck, A.R. Vasconcellos
and Luzzi R. \emph{ Markovian Kinetic Equations in a Nonequilibrium
Statistical Ensemble Formalism}, Phys. Rev. E \textbf{57}, 3637-3640
(1998).

\bibitem{vannucchi2009osc}F.S. Vannucchi, A.R. Vasconcellos, R.
Luzzi, \emph{Thermo-Statistical Theory of Kinetic and Relaxation Processes},
Int. J. Mod. Phys B \textbf{23}(27), 5283 (2009).

\bibitem{luzzi2000}R. Luzzi, A.R. Vasconcellos and J.G. Ramos,
\emph{Statistical Foundations of Irreversible Thermodynamics} (Teubner-BertelmannSpringer,
Stuttgart, Germany, 2000).

\bibitem{luzzi2001}R. Luzzi, A.R. Vasconcellos and J.G. Ramos,
\emph{Irreversible Thermodynamics in a Nonequilibrium Statistical
Ensemble Formalism}, Rivista del Nuovo Cimento \textbf{24}(3), 1-70,
(2001).

\bibitem{feynman1972}R. Feynman, \emph{Statistical Mechanics} (Benjamin,
Reading, USA, 1972).

\bibitem{fano1957}U. Fano, \emph{Description of States in Quantum
Mechanics by Density Matrix and Operator Techniques}, Rev. Mod. Phys.
\textbf{29}, 74-93 (1957).

\bibitem{balescu1975}R. Balescu, \emph{Equilibrium and Nonequilibrium
Statistical Mechanics} (Wiley Interscience, New York, USA, 1975).

\bibitem{hugenholz}N.M. Hugenholtz, Perturbation Theory of Large
Quantum Systems, in \emph{Quantum Theory of Many-Particle Systems,}
edited by L. van Hove, N.M. Hugenholtz (Benjamin, New York, USA,
1961), and Quantum Theory of Many-Body System, in \emph{Many-Body
Problems}, edited by W.E. Perry \emph{et al}. (Benjamin, New York,
USA, 1969).

\bibitem{zubarev?}D.N. Zubarev and M.Yu. Novikov, \emph{Renormalized
kinetic equations for a system with weak interaction and for a low-density
gas} Theor. Math. Phys. \textbf{19}(2), 480-490 (1974).

\bibitem{livshits1972}A.M. Livshits, \emph{Participation of Coherent
Phonons in Biological Processes}, Biofizika \textbf{17}(4), 694-695
(1972).

\bibitem{mori1965}H. Mori, \emph{Transport, Collective Motion, and
Brownian Motion}, Prog. Theor. Phys. (Japan) \textbf{33}, 423-455
(1965).

\bibitem{landsberg1981}P.T. Landsberg, \emph{Photons at Non-zero
Chemical Potential}, J. Phys. C \textbf{14}, L1025-L1027 (1981).

\bibitem{luzzi1997}R. Luzzi, A.R. Vasconcellos, J. Casas-Vazquez,
D. Jou, \emph{Thermodynamic Variables in the Context of a Nonequilibrium
Statistical Ensemble Approach}, J. Chem. Phys. \textbf{107}, 7383
(1997).

\bibitem{casas-vazquez2003}J. Casas-Vazquez, D. Jou, \emph{Temperature
in Non-equilibrium States: a Review of Open Problems and Current Proposals},
Rep. Prog. Phys. \textbf{66}, 1937-2023 (2003).

\bibitem{kim1990}D. Kim, P.Y. Yu, \emph{Phonon temperature overshoot
in GaAs excited by subpicosecond laser pulses}, Phys. Rev. Lett. \textbf{64},
946-949 (1990).

\bibitem{algarte1992}A.C. Algarte, A.R. Vasconcellos, R. Luzzi, \emph{Kinetics
of Hot Elementary Excitations in Photoexcited Polar Semiconductors},
Phys. Stat. Sol. (b) \textbf{173}, 487-514 (1992).

\bibitem{demidov2008b}V.E. Demidov, O. Dzyapko, S.O. Demokritov,
G.A. Melkov, A.N. Slavin, \emph{Observation of Spontaneous Coherence
in Bose-Einstein Condensate of Magnons}, Phys. Rev. Lett. \textbf{100},
047205 (2008).

\bibitem{walyazek1971}K. Walyazek, D.N. Zubarev, A.L. Kuzemskii,
\emph{Schr\"{o}dinger-type equation with damping for a dynamical system
in a thermal bath}, Theor. Math. Phys. \textbf{5}(2), 1150-1158 (1971).

\bibitem{scott1973}A.C. Scott, F.Y. Chu, A.L. McLaughlin, \emph{The
soliton: A new concept in applied science}, Proc. IEEE \textbf{61},
1443-1483 (1973).

\bibitem{davydov1980}A.S. Davydov, in \emph{Solitons}, R.K. Bullough,
P.J. Coudrey, Eds. (Springer, Berlin, Germany, 1980).

\bibitem{vasconcellos1998}A.R. Vasconcellos, M.V. Mesquita, R. Luzzi,
\emph{ Statistical Thermodynamic Approach to Vibrational Solitary
Waves in Acetanilide}, Phys. Rev. Lett \textbf{80}, 2008-2011 (1998).

\bibitem{vasconcellos1993}A.R. Vasconcellos, R. Luzzi, \emph{Vanishing
thermal damping of Davydov's solitons}, Phys. Rev. E \textbf{48},
2246-2249 (1993).

\bibitem{rezende2006}S. Rezende, F.M. Aguiar, A. Azevedo, \emph{Magnon
excitation by spin-polarized direct currents in magnetic nanostructures},
Phys. Rev. B \textbf{73}, 094402 (2006).

\bibitem{harris1963} A.B. Harris, \emph{Spin-wave Spectra of Yttrium
and Gadolinium Iron Garnet}, Phys. Rev. \textbf{132}(6), 2398-2409
(1963). 

\bibitem{cherepanov1993} V. Cherepanov, I. Kolokolo, V. L'vov, \emph{The
saga of YIG: Spectra, thermodynamics, interaction and relaxation of
magnons in a complex magnet}, Physics Reports \textbf{229}(3), 81-144
(1993).

\bibitem{kreisel2009} A. Kreisel et al, \emph{Microscopic spin-wave
theory for yttrium-iron garnet films}, Eur. Phys. J. B \textbf{71},
59-68 (2009). 

\bibitem{landaulifshitz_QE} V. B. Berestetskii, E. M. Lifshitz, L.
P. Pitaevskii, \emph{Quantum Eletrodynamics} (Butterworth Heinemann,
Amsterdam, The Netherlands, 1982), Course of Theoretical Physics,
v.4.
\end{thebibliography}
\end{document}